\documentclass[fleqn,usenatbib]{mnras}

\usepackage{newtxtext,newtxmath}

\usepackage[T1]{fontenc}
\usepackage{ae,aecompl}

\usepackage{graphicx}	
\usepackage{amsmath}	
\usepackage{amssymb}	

\newcommand{\blue}[1]{{\textcolor{blue}{#1}}}

\title[Halo assembly in misaligned galaxies]{SDSS-IV MaNGA: Signatures of halo assembly in kinematically misaligned galaxies}

\author[C. Duckworth et al.]{
\parbox{\textwidth}{Christopher Duckworth,$^{1}$\thanks{E-mail: cd201@st-andrews.ac.uk}
Rita Tojeiro,$^{1}$
Katarina Kraljic,$^{2}$
Mario A. Sgr\'o,$^{3,1}$ \\
Vivienne Wild,$^{1}$
Anne-Marie Weijmans,$^{1}$
Ivan Lacerna,$^{4,5}$
Niv Drory$^{6}$
\\
}\vspace{3mm}\\
{}$^{1}$School of Physics and Astronomy, University of St Andrews, North Haugh, St Andrews, KY16 9SS, UK\\
{}$^{2}$Institute for Astronomy, University of Edinburgh, Royal Observatory, Blackford Hill, Edinburgh EH9 3HJ, UK\\
{}$^{3}$Instituto de Astronom\'ia Te\'orica y Experimental, Observatorio Astron\'omico de C\'ordoba, Laprida 854, C\'ordoba, X500BGR, Argentina\\
{}$^{4}$Instituto de Astronom\'ia, Universidad Cat\'olica del Norte, Av. Angamos 0610, Antofagasta, Chile\\
{}$^{5}$Instituto Milenio de Astrof\'isica, Av. Vicu\~na Mackenna 4860, Macul, Santiago, Chile\\
{}$^{6}$McDonald Observatory, The University of Texas at Austin, 1 University Station, Austin, TX 78712, USA\\
}

\date{Accepted XXX. Received YYY; in original form ZZZ}

\pubyear{2018}

\begin{document}
\label{firstpage}
\pagerange{\pageref{firstpage}--\pageref{lastpage}}
\maketitle

\begin{abstract}
We investigate the relationship of kinematically misaligned galaxies with their large-scale environment, in the context of halo assembly bias. 
According to numerical simulations, halo age at fixed halo mass is intrinsically linked to the large-scale tidal environment created by the cosmic web. We investigate the relationship between distances to various cosmic web features and present-time gas accretion rate. We select a sub-sample of $\sim$900 central galaxies from the MaNGA survey with defined global position angles (PA; angle at which velocity change is greatest) for their stellar and H$\alpha$ gas components up to a minimum of 1.5 effective radii ($R_e$). We split the sample by misalignment between the gas and stars as defined by the difference in their PA. For each central galaxy we find its distance to nodes and filaments within the cosmic web, and estimate the host halo's age using the central stellar mass to total halo mass ratio $M_{*}/M_{h}$. We also construct halo occupation distributions using a background subtraction technique for galaxy groups split using the central galaxy's kinematic misalignment. We find, at fixed halo mass, no statistical difference in these properties between our kinematically aligned and misaligned galaxies. We suggest that the lack of correlation could be indicative of cooling flows from the hot halo playing a far larger role than `cold mode' accretion from the cosmic web or a demonstration that the spatial extent of current large-scale integral field unit (IFU) surveys hold little information about large-scale environment extractable through this method.
\end{abstract}




\section{Introduction}
In the $\Lambda$ cold dark matter ($\Lambda$CDM) paradigm, galaxies form in the gravitational potential wells of dark matter haloes \citep{white1978}. Within this framework, structure grows hierarchically, and dark matter haloes today are the direct result of bottom-up assembly, with smaller haloes forming first and merging to then form larger ones. This process can be explained approximately by the excursion-set formalism which tracks the linear growth of primordial over-densities before spherically collapsing in the non-linear regime \citep{press1974,bond1991}. In the most basic form, environment is neglected and the assembly history of the halo is entirely dictated by its mass. This exclusive dependence on mass underlines the widely used halo occupation distribution \citep[HOD; e.g.][]{jing1998,peacock2000} modelling and various types of abundance matching \citep[e.g.][]{kravtsov2004,conroy2006}, which require galaxy clustering to be driven solely by halo mass \citep[e.g.][]{mo1996,sheth1999}. 

Parallel to the successes of HOD modelling, N-body simulations have fast converged on the fact that, at fixed halo mass, haloes which have formed at different times cluster differently \citep[e.g.][]{gao2005,wechsler2006,croton2007,wang2011}. This effect has been called \textit{halo assembly bias} and quantifies any physical quantity that determines halo clustering beyond halo mass. Halo formation time is most commonly considered for halo assembly bias. However both halo spin and concentration have been motivated to affect formation and assembly \citep[e.g.][]{lacerna2012,lehmann2017}. Attempts to understand the origin of assembly bias at low halo mass have come from the large-scale tidal environment in which a given halo resides. \citet{hahn2009} found a systematic trend between halo formation time and the large-scale tidal force strength, derived from the geometric environment, at fixed halo mass. This effect is seen most strongly in low-mass haloes, arising from suppression of their growth when they reside within the vicinity of a much larger halo. This large halo acts to stop accretion on the low-mass halo in over-dense regions, effectively boosting the clustering of older low-mass haloes, compared to haloes of the same mass residing in under-dense regions that are less affected by tidal fields. \citet{ZOMGI} explore this phenomenon in the context of the cosmic web. Low-mass haloes residing within large filaments can often see their accretion `stalled' and hence will cease mass assembly earlier as matter flows preferentially along the filament to its densest points (nodes). Conversely low-mass haloes at the convergence point of multiple smaller filaments will have continued isotropic accretion resulting in longer continued mass growth and more recent formation. \citep[See][for a theoretical approach]{musso2018}.

Galaxies are our primary resource in probing the spatial distribution of dark matter. Their formation and subsequent evolution is tied to the assembly history of their host halo, however, determining the exact link is difficult. The observational counterpart of assembly bias is therefore tricky to isolate and as such is rightfully still under debate. 

Early studies, however, have demonstrated observations of halo assembly bias. For example, \citet{tojeiro2017} compare a halo age proxy with respect to large-scale tidal environment defined in the Galaxy And Mass Assembly \citep[GAMA;][]{driver2009, driver2011} survey. They quantify tidal strength using the geometric classification of \citet{eardley2015} to characterise regions into geometric voids, sheets, filaments and knots corresponding to zero, one, two and three dimensions of collapse respectively. They find that low-mass haloes ($\lesssim 10^{12.5} M_{\odot}$) show a steadily increasing ratio of central galaxy stellar mass to total halo mass, corresponding to increasing halo age in regions of increasing tidal force strength (i.e. going from voids to knots). They find a tentative reversal of this trend for high-mass haloes ($\gtrsim 10^{13.2} M_{\odot}$). \citep[See][who explicitly look for changes in halo to stellar mass ratio with geometric environment using stacked lensing profiles, but find no significant changes when averaging over halo mass.]{brouwer2016}.

The tidal field can also be described in a topological sense with respect to the cosmic web. \citet{kraljic2018} provide an investigation in the GAMA survey through identification of the cosmic web, using the Discrete Persistent Structure Extractor code \citep[DisPerSE;][]{sousbie2011a,sousbie2011b}. They estimate distances to nodes, filaments and walls as a function of galaxy properties such as $u - r$ colour, specific star formation rate (sSFR) and stellar mass. They find distinct gradients with more massive, redder (passive) galaxies residing closer to nodes, filaments and walls, indicative of mass dependent clustering. Additionally, at fixed stellar mass, both star formation rate (SFR) decreases and colour reddens for galaxies closer to both nodes and filaments. Assuming the flow of baryonic accretion follows that of dark matter, this observation is consistent with the `stalling' of haloes due to tidal environment. 

In this paper we look at stellar and gas kinematics as a potential indicator for recent halo accretion and hence late-time assembly. Galaxies are subject to a host of external processes which both replenish and disturb their gas component. Accreted gas could originate from the cold filamentary flows of the cosmic web (which also boosts the prevalence of minor mergers), the cooling flows of the surrounding hot halo or conversely be a consequence of multiple major mergers with neighbouring galaxies. 

Integral Field Spectroscopic (IFS) surveys provide spatially resolved properties of galaxies enabling detailed identification of external influence. Previous work has focussed on the gas content in early-type galaxies (ETGs) and the fraction of which that are significantly kinematically misaligned (i.e. global position angle between rotational directions of gas and stars is greater than 30$^{\circ}$) indicating external influence. \citet{davis2011a} identify that approximately 36\% of fast-rotating ETGs are kinematically misaligned within the volume-limited sample of ATLAS\textsuperscript{3D} \citep{atlas3d}, setting a lower limit on the importance of externally acquired gas. 
This can also be used to constrain the time-scales of misalignment. \citet{davis2016} utilise a toy model to propose that misaligned gas could relax gradually over time-scales of 1-5 Gyr. A faster time-scale of relaxation would require merger rates of $\approx 5$ Gyr\textsuperscript{-1} and hence is disfavoured. The interplay between the strength and persistence of the gas in-flow and the re-aligning torque of the stellar component dictates the exact time-scale of misalignment for an individual galaxy. The strength of a galaxy's stellar torque scales as a function of radius, with the central component of a galaxy re-aligning on a quicker time-scale than the outer regions. 

The persistence of misalignment has also been considered in numerical simulations. \citet{vdvoort2015} consider the evolution of a misaligned gas disc formed from a merger which removes most of the original disc. During re-accretion of the cold gas, the misaligned disc persists for approximately 2 Gyr before the gas-star rotation angle falls below 20$^{\circ}$. The sustainment of this misalignment is due to continued gas accretion for approximately 1.5 Gyr before its rate falls and the gas can realign with the stellar component on approximately six dynamical time-scales. 

At the present-time, there are two large-scale IFS surveys which could provide identification of kinematic misalignment with respect to halo assembly or environment for a statistically significant sample. The Sydney-AAO Multi-object Integral field spectrograph survey \citep[SAMI;][]{croom2012,bryant2015} will map $\sim$3400 galaxies in the local universe ($z < 0.12$) across a large range of environments in the GAMA footprint. In parallel, the Mapping Nearby Galaxies at Apache Point \citep[MaNGA;][]{bundy2015,blanton2017} survey will map $\sim$10000 galaxies up to $z = 0.15$, with the aim of creating a sample with near flat number density distribution in absolute $i$-band magnitude and stellar mass. 

The impact of environment on kinematically misaligned galaxies has been previously studied in MaNGA. Using a sample of 66 misaligned galaxies, \citet{jin2016} found that the fraction of misalignment varies with galaxy properties such as stellar mass and sSFR. Regardless of sSFR, they find that kinematically misaligned galaxies are typically more isolated. Could these correspond to the later forming haloes? 


We explore whether position with respect to filamentary structures identified in the cosmic web, halo age and estimated group occupancy correlate with more recent accretion observed on the central galaxy of the group. A younger halo corresponds, by definition, to more recent dark matter accretion and a richer recent merger history. We explore the idea that low-mass haloes have their cold flow accretion halted due to large-scale tidal environment as indicated by the halo age or vicinity to cosmic web features. We also discuss the prevalence of gas cooling from the hot halo in interpretation of our results and consider the ability of the global position angle to identify gas accretion.

The paper is structured as follows. Section 2 describes the MaNGA survey and the corresponding data we use in this work. Section 3 describes our sample selection from MaNGA and cross matching to various environmental parameters. Section 4 motivates our three different measures of assembly history and environment: distance to various cosmic web features (Section 5), halo age (Section 6), and halo occupation distribution modelling (Section 7). In Section 8, we discuss assumptions in this work and their impact on our analysis, before concluding in Section 9. 
Throughout this paper we use the following cosmological parameters: $H_{0} = 67.5$ km s$^{-1}$ Mpc$^{-1}$, $\Omega_{m} = 0.31$, $\Omega_{\Lambda} = 0.69$ \citep{planck2016}. 

\section{Data}
\subsection{The MaNGA survey} \label{MaNGA_survey}
The MaNGA survey is designed to investigate the internal structure of an unprecedented $\sim$10000 galaxies in the nearby Universe. MaNGA is one of three programs in the fourth generation of the Sloan Digital Sky Survey (SDSS-IV) which enables detailed kinematics through integral field unit (IFU) spectroscopy. Using the 2.5-metre telescope at the Apache Point Observatory \citep{gunn2006} along with the two channel BOSS spectrographs \citep{smee2013} and the MaNGA IFUs \citep{drory2015}, MaNGA provides spatial resolution on kpc scales (2'' diameter fibres) while covering 3600-10300$\mathring{A}$ in wavelength with a resolving power of R$\sim$2000. 

\begin{figure}
	\includegraphics[width=\linewidth]{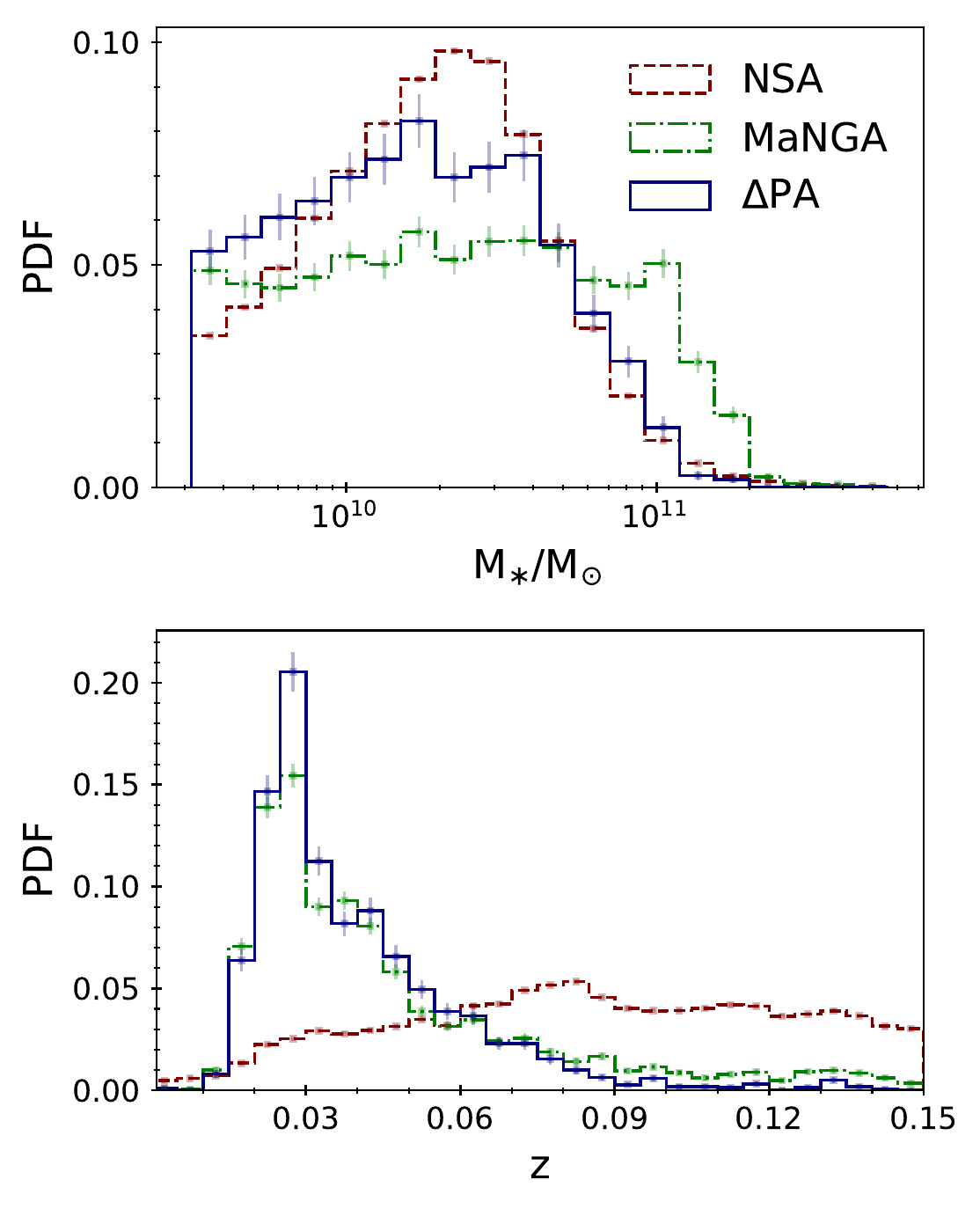}
    \caption{Relative frequency distributions of stellar mass and redshift for the NSA target catalogue (brown dashed line), MaNGA MPL-6 (green dot-dashed line) and our $\Delta$PA sub-sample (blue solid line). The figure is cut at $z=0.15$ representing the extent of MaNGA targets. Each histogram is given with Poisson errors on each bin.}
    \label{fig:samp_cons}
\end{figure}

By design, the complete sample is unbiased towards morphology, inclination and colour and provides a near flat distribution in stellar mass. This is enabled by three major observation subsets: the Primary sample, the Secondary sample and the Colour-Enhanced supplement. All sub-samples observe galaxies to a minimum of $\sim 1.5$ effective radii ($R_{e}$) with the Secondary sample increasing this minimum to $\sim 2.5 R_{e}$. The Colour-Enhanced supplement fills in gaps of the colour-magnitude diagram leading to an approximately flat distribution of stellar mass. A full description of the MaNGA observing strategy is given in \citet{law2015obs,yan2016obs}. Figure \ref{fig:samp_cons} shows the distribution of stellar mass and redshift of the sixth MaNGA Product Launch (MPL-6) with comparison to the NASA-Sloan Atlas (NSA) catalogue from which MaNGA is targeted and our $\Delta$PA defined sub-sample outlined in \S\ref{sec:samp_sec}. While this selection is naturally not volume limited, it is a simple exercise to correct for since every galaxy in the redshift range is known \citep{wake2017}. 

MaNGA observations are covered plate by plate, employing a dithered pattern for each galaxy corresponding to one of the 17 fibre-bundles of 5 distinct sizes. Data is provided by the MaNGA data reduction pipeline as described in \citet{law2016drp,yan2016spec}. Any incomplete data release of MaNGA should therefore be unbiased with respect to IFU sizes and hence a reasonable representation of the final sample scheduled to be complete in 2020. In this work, we use galaxies from MPL-6. This corresponds to 4633 unique galaxies which will go public with slightly different data reduction as part of DR15 in December 2018.

\subsection{Velocity maps}
All stellar and gas velocity maps are output from the MaNGA Data Analysis Pipeline (DAP; Westfall et al. in prep). A complete discussion will be given in Westfall et al. (in prep), however for the time being we will summarise the key points here. 

Stellar kinematics are derived using the Penalised Pixel-Fitting (pPXF) method \citep{cappellari2004,cappellari2017}. This extracts the line of sight velocity dispersion and then fits the absorption-line spectra from a set of 49 clusters of stellar spectra from the MILES stellar library \citep{sanchez2006,falcon2011}. Before extraction of the mean stellar velocity, the spectra are spatially Voronoi binned to $g$-band \textit{S/N} $\sim$ 10, excluding any individual spectrum with a $g$-band \textit{S/N} < 1 \citep{cappellari2003}. This approach is geared towards stellar kinematics as the spatial binning is applied to the continuum \textit{S/N}, however, we note that unbinned and Voronoi binned velocity maps produce similar results. 

Gas velocity fields are extracted through fitting a Gaussian to the H$\alpha$-6564 emission line, relative to the input redshift for the galaxy. This velocity is representative for all ionized gas, since the parameters for each Gaussian fit to each emission line are tied during the fitting process. These velocities are also binned spatially by the Voronoi bins of the stellar continuum.

\subsection{Defining misalignment} \label{def_mis}
\begin{figure*}
	\includegraphics[width=\linewidth]{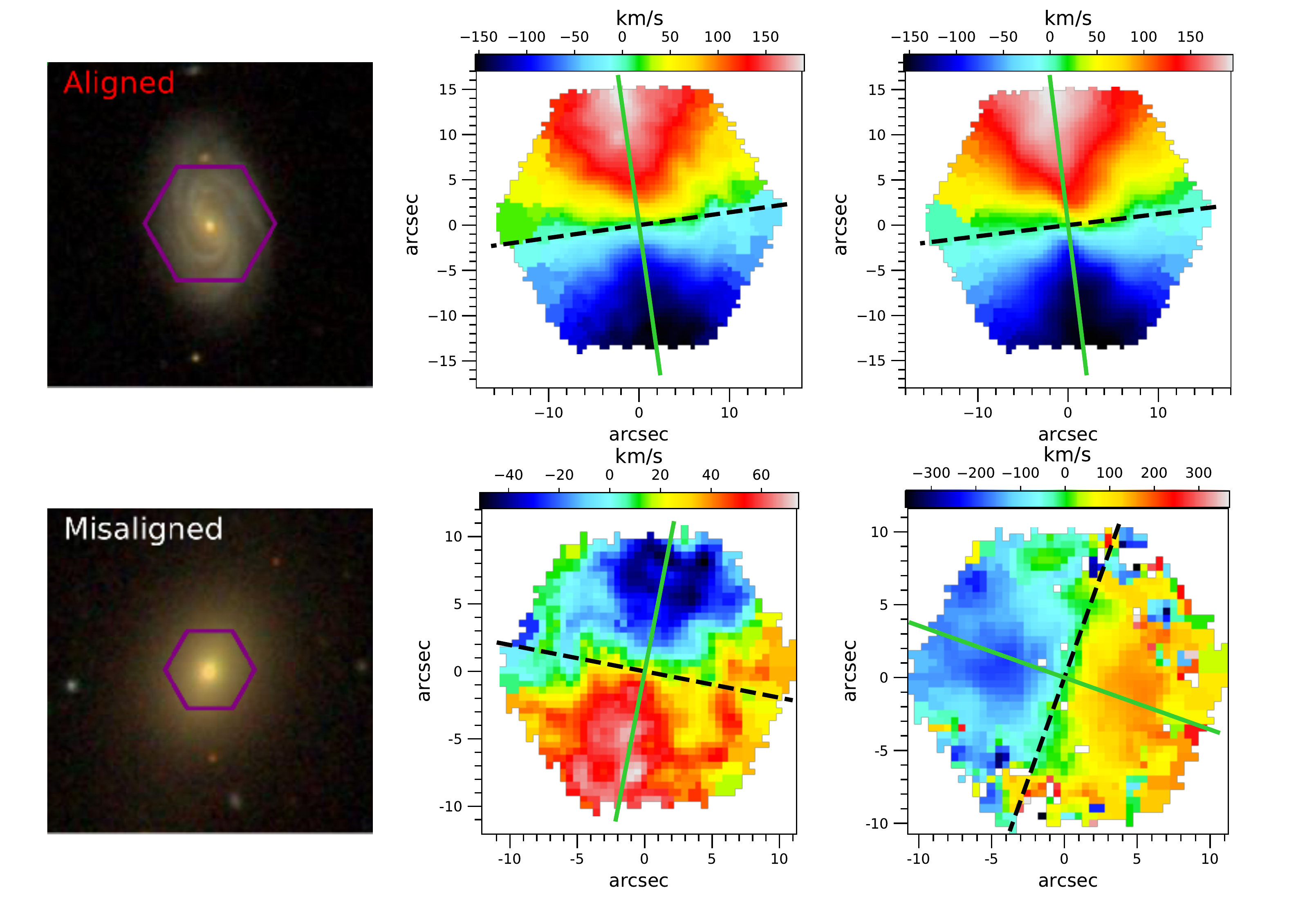}
    \caption{Examples of a kinematically aligned (top) and misaligned (bottom) galaxy defined by $\Delta$PA. From left to right, the panels show (i) the original SDSS cutout of surrounding field with the MaNGA IFU footprint overlaid in purple, (ii) stellar velocity field and (iii) $H\alpha$ (gas) velocity field. The velocity fields are marked by a defined PA (green solid line) and axis of rotation (black dashed line). The galaxy in the bottom row is misaligned due to it having $\Delta$PA > 30$^{\circ}$. The colour-bars represent the velocity fields in km/s and the galaxies are orientated so that up corresponds to North and right to East.}
    \label{fig:cutout_wIFU}
\end{figure*}

To identify accreting galaxies, we estimate the two dimensional global position angle (PA) of the stellar and H$\alpha$ gas velocity fields using the \texttt{FIT\_KINEMATIC\_PA} routine outlined in \citet{krajnovic2006}. By default this finds the angle corresponding to the bisecting line which has greatest velocity change along it (i.e. the angle of peak rotational velocity). We choose this angle to be found from sampling at 180 equally spaced steps. This is measured counter-clockwise from the north axis, however, it does not discriminate between the blueshifted and redshifted side since it is only defined up to 180$^{\circ}$. As a result, velocity fields with a difference of 180$^{\circ}$ PA would appear to be aligned. To solve this we identify the direction of rotation and re-assign a consistent PA: defined as the axis of rotation approximately 90$^{\circ}$ clockwise from the blueshifted side. This angle now spans 360$^{\circ}$ allowing an automatic detection of misaligned gas and stellar components. The offset angle between kinematic components is defined as: 
\begin{equation} \label{eq:delPA}
\Delta PA = |PA_{stellar} - PA_{H\alpha}|. 
\end{equation}
We define galaxies with $\Delta$PA > 30$^{\circ}$ to be significantly kinematically misaligned. An example of an aligned and a misaligned galaxy is shown in Figure \ref{fig:cutout_wIFU}. 

To improve the reliability of the PA fit, we apply a few additional filters to the velocity fields. While foreground stars are flagged within observations, background/small neighbouring galaxies can remain within the IFU footprint. This is a problem for fitting a global position angle since it naturally interprets other material as part of the target galaxy's observation and interpolates between the regions. We remove all disconnected regions smaller than $10\%$ of the target galaxy's footprint. In addition we sigma clip the velocity field and remove all spectral pixels (spaxels) above a $3\sigma$ threshold.

We construct two component model velocity maps for each stellar and gas component of every MPL-6 observation in order to estimate typical errors on $\Delta$PA; see Appendix A for a full discussion. We find that \texttt{FIT\_KINEMATIC\_PA} gives a typical combined (stellar and gas) mean error of $1.3^{\circ}$, for maps with simplistic motions.

Assuming gas that has cooled from stellar mass loss to be kinematically aligned with the original stars, misalignment is indicative of external gas origin \citep[see][]{sarzi2006,davis2011a}. Active Galactic Nuclei (AGN) may also be assumed to not drive significant misalignment. A study of 62 AGN galaxies in MaNGA found no difference in the distribution of $\Delta$PA with respect to an inactive control (Ilha et al. in prep). 

Our choice to take $\Delta$PA > 30$^{\circ}$ as significantly kinematically misaligned is a conservative selection to ensure we are selecting galaxies undergoing external interaction. There is evidence to suggest that accretion drives misalignment past $\Delta$PA = 30$^{\circ}$. \citet{lagos2015} found that using solely galaxy mergers as the source for misaligned cold gas only predicts 2\% of ETGs to have $\Delta$PA > 30$^{\circ}$ using GALFORM, in comparison with the misaligned field ETG fraction found in ATLAS\textsuperscript{3D} of 42 $\pm$ 6\%  \citep{davis2011a}. This puts a lower level of importance on gas accretion. Our choice can be justified as follows. Firstly, we are using ionized gas as a proxy for the distribution and accretion of cold gas. \citet{davis2011a} find that the typical difference between the PAs of cold gas (CO) and ionized gas can be described by a Gaussian distribution centred on 0 with a standard deviation of 15$^{\circ}$ for 38 CO bright galaxies in ATLAS\textsuperscript{3D}. While indicating ionized gas is a reasonable estimator for cold gas, splitting $\Delta$PA = 30$^{\circ}$ accounts for the scatter in this relationship. Secondly, this should avoid spurious misalignments arising from errors in the fit of $\Delta$PA. While our model errors are low, they are likely an underestimation since they do not include more complex motions. However, selecting a lower split in $\Delta$PA would only be affected by the increased likelihood of internal processes being dominant, rather than the inaccuracy of fitting. Any threshold in $\Delta$PA becomes a trade off between sample size and contamination probability. Altering our cut in $\Delta$PA to be 20$^{\circ}$, 40$^{\circ}$ or similar does not change any of the conclusions drawn in this work.

\subsection{Stellar mass and halo mass definitions} \label{mass}
We use stellar masses estimated in the New-York University Value Added Catalogue from the K-correct routine \citep[NYU-VAC;][]{blanton2005}. To analyse the incident effects onto central galaxies as a function of environment, we require both group identification and estimations of the total halo mass. \citet{yang2007} (Y07 hence-forth) present an adaptive group finding algorithm based on the NYU-VAC to assign galaxies to haloes and then estimate and revise group properties through iteration. We will summarise the basic steps here. Potential group centres are first found through a friend-of-friends algorithm with small linking lengths in redshift space. All galaxies not currently linked are also considered as potential centres. For each tentative group, the combined luminosity of all group members with 
\begin{equation}
^{0.1}M_r - 5log(h) \leq -19.5
\end{equation}
are found, where $^{0.1}M_r$ is the absolute magnitude estimated from the NYU-VAC K-corrected to $z=0.1$. All galaxies within SDSS data-release 7 (DR7, all spectroscopic galaxies) below $z=0.09$ meet this criteria and hence the combined luminosity can be estimated directly. A corrective factor is applied for groups above this redshift. This total luminosity is then used to assign halo mass and various other group properties, which in turn refines the group identification following an iterative process until conversion. 

We remove all groups with $f_{edge} < 0.6$ as recommended by Y07 corresponding to groups on the survey borders. \citet{yang2009} provide a conservative estimate on the minimum halo mass for groups that would be expected to be complete as a function of redshift. This work primarily focusses on low-mass haloes, ($M_h \lesssim 10^{12.5}$), however many of these groups will be incomplete at the redshift range selected. Y07 note however that any potential scatter arising from the incompleteness correction should be minimal in comparison with the scatter between group luminosity and assigned halo mass. 

The use of a group catalogue brings important considerations. In low-mass groups, the multiplicity of each group is low, and the luminosity (or stellar mass) of the central galaxy, will largely determine the mass of the group, potentially leading to an under-estimation of the scatter in the stellar-mass to halo-mass relation (see e.g. \citealt{campbell2015, reddick2013}). On the other hand, the fraction of groups with at least one satellite galaxy that is more massive than the central, increases steeply with halo mass (see e.g. \citealt{reddick2013}), leading to artificially larger scatter and an increased likelihood of central misclassification with increasing halo mass. \citet{campbell2015} demonstrate that group finder inferred measurements tend to equalise the properties of distinct galaxy sub-populations, however in general it is possible to recover meaningful physical correlations for average properties as a function of stellar and halo mass.  

\begin{figure}
	\includegraphics[width=\linewidth]{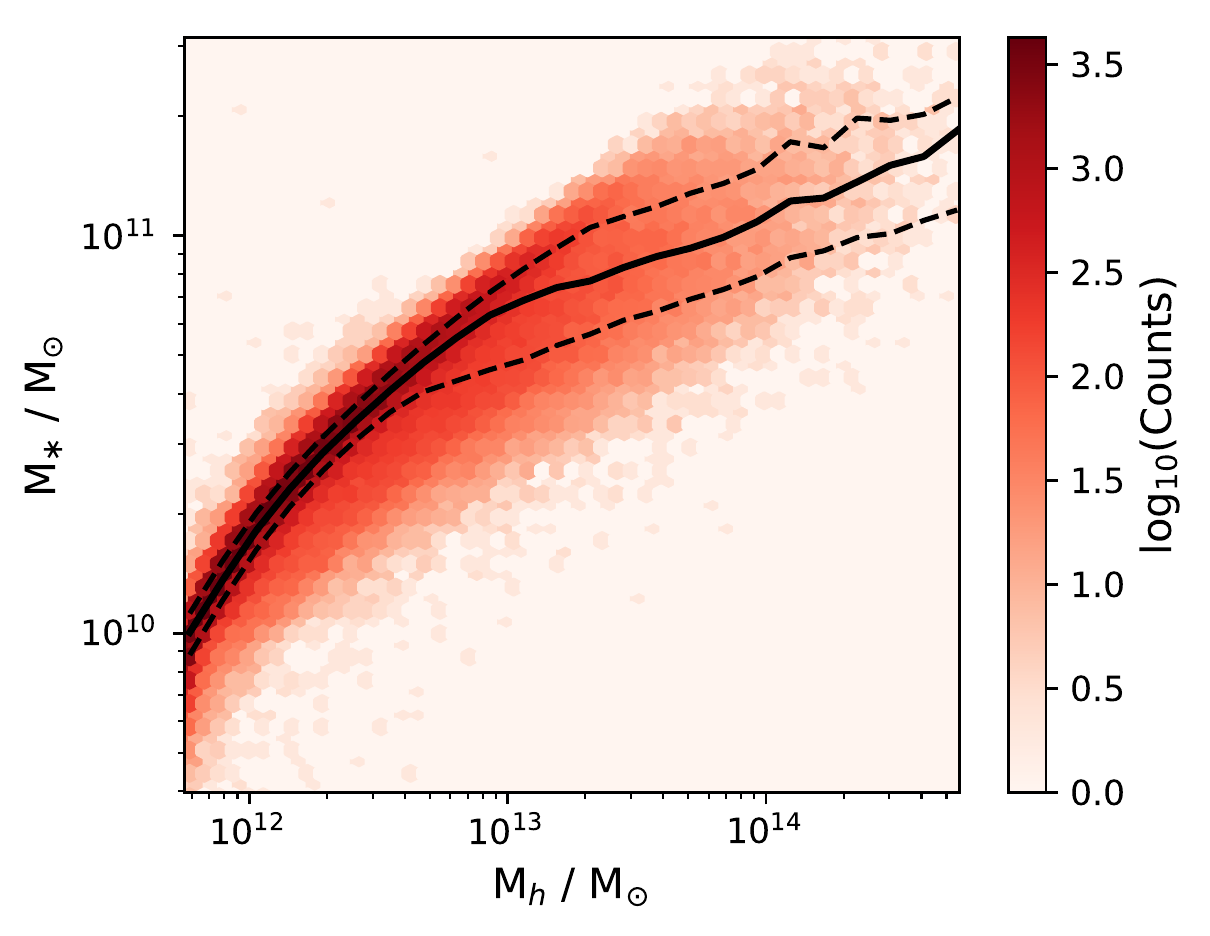}
    \caption{Hexagonal density plot of the relationship between the stellar mass  and the halo mass for central galaxies within Y07. The number counts are scaled logarithmically. The black lines show the median (solid) and the 20th and 80th quantiles (dashed) of the stellar mass in bins of halo mass.}
    \label{fig:stel_halo_dist}
\end{figure}

We select only galaxies classified as centrals corresponding to the most massive galaxy within each group identified by Y07. The relationship between the central stellar mass and halo mass for galaxies within Y07 is shown in Figure \ref{fig:stel_halo_dist}. The increase of scatter in this relationship with increasing halo mass is likely explained by the limitations of group catalogues mentioned above. 

\section{Sample selection} \label{sec:samp_sec}
While the complete MaNGA sample may be unbiased to morphology, we must proceed with care when selecting a sub-sample usable for kinematic disturbance. To remove spurious PA fits we eyeball the entire MPL-6 sample and remove all galaxies which have a largely incomplete velocity field, poor or biased PA fit or are virtually face-on so that little or no rotational component is along the line of sight. This removes approximately half of MPL-6 observations. The majority of galaxies removed have largely incomplete gas velocity fields and for that reason our analysis naturally excludes gas-poor and slowly rotating elliptical galaxies. 

Recent studies have found that the fraction of slow rotators increases steeply with stellar mass and has a weak dependence on environment once this is controlled for \citep[e.g.][]{greene2018,lagos2018}. We note that our natural exclusion of gas-poor high-mass galaxies is reflected in the stellar mass distribution of the $\Delta$PA defined sample in Figure \ref{fig:samp_cons}.

GalaxyZoo1 provides visually identified morphologies for a large sample of SDSS galaxies \citep{lintott2008}. Morphology is identified by having over 80\% of debiased classification votes in the same category (i.e. elliptical or spiral). The remainder of galaxies are marked as uncertain morphology. We compare the fraction of ellipticals in our $\Delta$PA sub-sample with MPL-6, as shown in Table \ref{tab:GZ}. GalaxyZoo1 only provides classifications for 3/4 of MPL-6 as reflected by the total classification numbers. We find the fraction of ETGs falls from 0.242 to 0.111, reaffirming our bias against slow-rotating high-mass ellipticals that our eye-balling tends to remove. 

If galaxies are truly morphologically transformed, then this should be reflected in their angular momentum. \citet{cortese2016} find that galaxies lie on a tight plane defined by stellar angular momentum ($j_{stars}$), S\'ersic index and stellar mass when excluding slow rotators in the SAMI galaxy Survey. This could indicate that fast rotating early-type and late-type galaxies are not two distinct populations but instead represent a continuum connecting pure-discs to bulge dominated systems \citep{cappellari2011}. This can be linked to simulation: \citet{lagos2017} use EAGLE \citep{EAGLE2015} to investigate the effects of galaxy mergers on the evolution of stellar specific angular momentum. They find that the gas content of a merger is the most important factor for dictating $j_{stars}$ for the remnant, ahead of both the mass ratio and spin/orbital orientation of the merger progenitors. An increasing rate of wet (gas-rich) mergers corresponds to decreasing stellar mass and increasing $j_{stars}$. Conversely dry (gas-poor) mergers are the most effective way of spinning down galaxies, with gas-poor counter-rotating progenitors creating the biggest decrease in $j_{stars}$. Following this narrative, it is fair to exclude slow rotators which could follow a different evolutionary track to a continuum of fast rotating galaxies in angular momentum phase space. 

An interesting result of the GalaxyZoo1 classification is that despite the $\Delta$PA defined galaxies in this analysis being predominately classified as LTGs, we find the majority of kinematically misaligned galaxies are ETGs. The ubiquity of misalignment in ETGs and lack there-of in LTGs is however a distinct question which could be explained by the relative relaxation time-scales of galaxies of different intrinsic angular momenta or the fractions of in-situ/ex-situ origin of gas. This could, however, be a natural result of disc formation and sustainment arising from cold flows of the cosmic web \citep{pichon2011}. If the disc is preferentially aligned with its larger surrounding structure then further directional accretion would be unlikely to create a kinematic misalignment. The low fraction of kinematically misaligned blue galaxies was first noted by \citet{chen2016} who explored a possible mechanism for their formation and characteristics. Selecting only early-type galaxies and comparing the misaligned sub-sample with a stellar mass weighted control does not fundamentally change any of the results presented here. While understanding how a morphology bias could impact any results presented, we leave its origin the focus of future work. 

\begin{table}
\begin{tabular}{|l|c|c|c|c|}
\hline
& Total & GZ1 & ETG & LTG \\ \hline
MPL-6 & 4614 & 3598 & 869 (0.242) & 1225 (0.340) \\
$\Delta$PA defined & 2272 & 1835 & 204 (0.111) & 1005 (0.548) \\
$\Delta$PA > 30$^{\circ}$ & 192 & 151 & 85 (0.556) & 9 (0.060)\\
Final sample & 925 & 812 & 136 (0.167) & 456 (0.561)
\end{tabular}
\caption{(Rows: top to bottom) All usable MPL-6 galaxies, all $\Delta$PA defined galaxies within MPL-6, those that are kinematically misaligned with $\Delta$PA > 30$^{\circ}$ and the final sample of central, $\Delta$PA defined galaxies used in this work. For each row, the total number of galaxies is given, along with those defined in GalaxyZoo1 (denoted GZ1 in table) and the total number of which that are classified into early-type (ETG) and late-type (LTG). The fractions of early-types and late-types are defined with respect to the total number of GalaxyZoo1 defined galaxies.}
\label{tab:GZ}
\end{table}

We are looking for accretion due to large-scale influence, so we remove all obvious on-going mergers through visual inspection of both the field photometry and IFU observations. We also identify target galaxies interacting with close pairs or neighbours. While this visual inspection should identify the majority of on-going major mergers, we note that our identifications are clearly subjective. We remove $\sim$50 galaxies, identified to be merging or interacting with a nearby neighbour. Table \ref{tab:interact}, at the end of the main text, provides all $\Delta$PA defined galaxies that were removed in this eyeballing. After matching to the Y07 group catalogue for halo mass we are left with 925 central galaxies which we use in this work. 

\section{Estimators of halo age and environment} 
In this work, $\Delta$PA of central galaxies will be correlated with various environment dependent parameters for the surrounding halo. Following the discussion of \citet{hahn2009}, the current accretion rate is correlated with the large-scale tidal environment. Low-mass haloes in the vicinity of large haloes or massive structures will have their accretion `stalled' as tidal forces overcome their ability to accrete \citep[see also;][]{wang2007,dalal2008,lacerna2011}. Accordingly, this would lead to a relatively earlier formation time and a lack of on-going accretion seen today. Conversely low-mass haloes in environments of low magnitude tidal forces will continue accreting, and could have a late halo assembly time and continued gas accretion onto the central galaxy today. In this paper we explore these ideas with the following tests:
\begin{enumerate}
\item Cosmic web classification (Section 5)
\begin{itemize}
\item Distances of haloes to filamentary structures are considered for galaxies split on $\Delta$PA. We test if low-mass haloes near filaments have their accretion `stalled' due to material preferentially flowing along the filament to more dense regions. 
\end{itemize}
\item Stellar to halo mass ratio  (Section 6)
\begin{itemize}
\item The stellar to halo mass ratio is used as a proxy for halo age and its correlation with $\Delta$PA is considered. We test if large-scale tidal forces can `stall' accretion onto low-mass haloes, seen as a low accretion rate today ($\Delta$PA < 30$^{\circ}$) on the central galaxy, indicating an earlier forming halo.
\end{itemize}
\item HOD modelling (Section 7)
\begin{itemize}
\item A halo occupation distribution is constructed for groups with aligned and misaligned central galaxies. Earlier forming haloes provide more time for centrals to form and satellites to merge. This would correspond to a decrease in the magnitude of the HOD, which we aim to isolate. 
\end{itemize}
\end{enumerate}
In each section we present both our method and results. 
\section{Cosmic web classification}
We explore the ability of a low-mass halo to accrete with respect to where it falls within the cosmic web. 
To characterise topological features, such as the filaments which could lead to the suppression of accretion, the Discrete Persistent Structure Extractor code \citep[DisPerSE;][]{sousbie2011a,sousbie2011b} is applied to a modified SDSS DR10 spectroscopic catalogue \citep{tempel2014}. 

\subsection{DisPerSE}

\begin{figure*}
	\includegraphics[width=\linewidth]{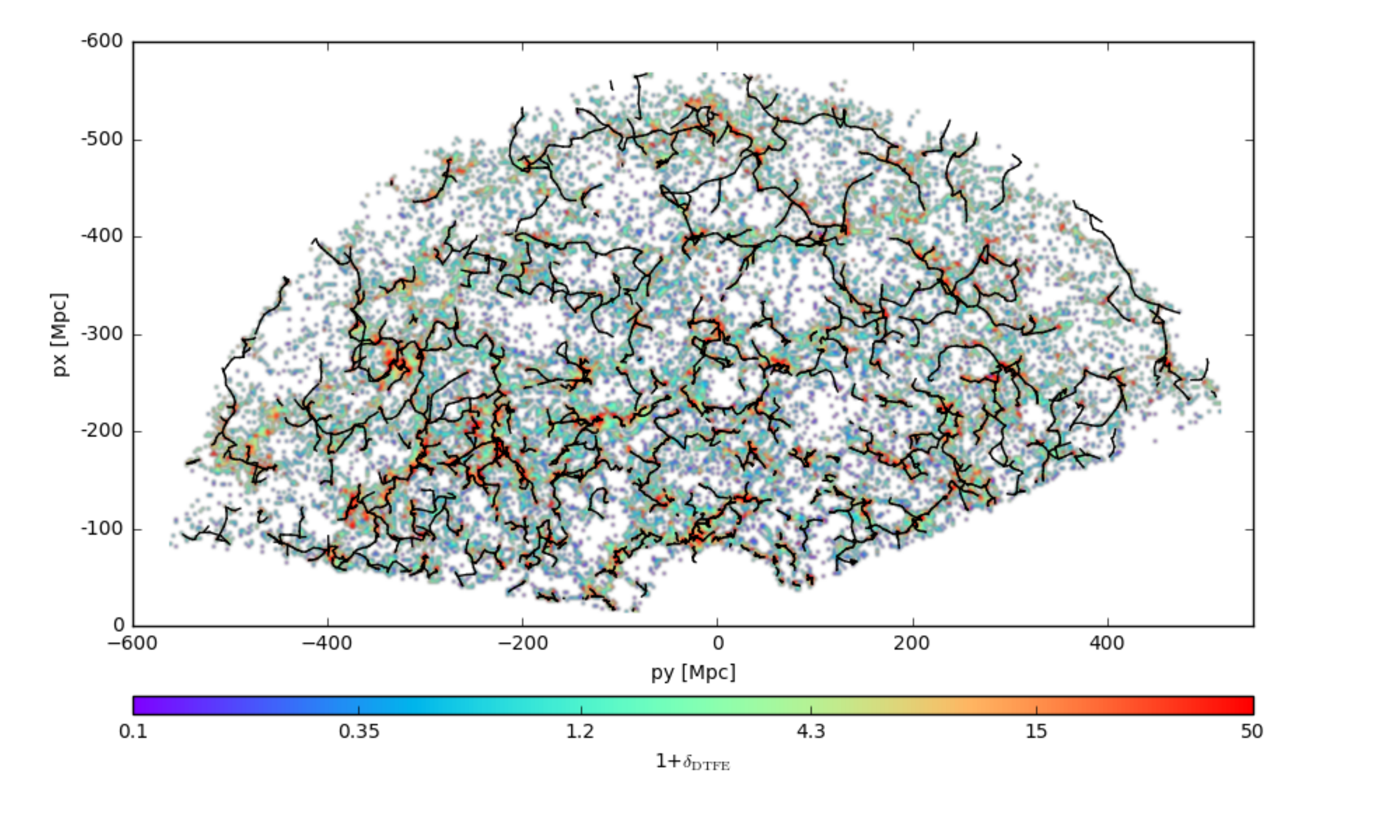}
    \caption{Illustration of the filamentary network (black lines) for a slice of the SDSS field ($0.02 \leq z \leq 0.15$; $ 27 \leq$ dec $\leq 33$) extracted using the DisPerSE code. Only filamentary structures which are seen to persist above the 5$\sigma$ threshold are shown, along with the density contrast of the galaxy population. The density contrast is estimated using the small-scale DTFE estimator (see text).}
    \label{fig:disperse}
\end{figure*}

DisPerSE is a geometric three-dimensional ridge extractor that applies directly to point-like distributions, making it particularly well adapted for astrophysical applications, as demonstrated by its previous application to various large galaxy surveys, such as SDSS \citep{sousbie2011b}, GAMA \citep{kraljic2018} or VIPERS \citep{malavasi2017}. 

It is based on discrete Morse and persistence theories, allowing for a scale and parameter-free coherent identification of the 3D structures of the cosmic web as dictated by the large-scale topology. 

In a nutshell, discrete homology is used to build the so-called Morse-Smale complex on the point-like tracers. This geometrical segmentation of space defines distinct regions called ascending manifolds that are identified as individual morphological components of the cosmic web; i.e. ascending manifolds of dimension 3, 2, 1 and 0 as voids, walls, filaments and nodes of the cosmic web, respectively. 

In addition to its ability to work with sparsely sampled data sets and assuming nothing about the geometry or homogeneity of the survey, retained structures can be selected on the basis of their significance compared to shot noise. DisPerSE hence allows to trace precisely the locus of filaments, walls and voids using the so called persistence ratio, a measure of the significance of the topological connection between individual pairs of critical points, mimicking thus an adaptive smoothing depending on the local level of noise. In practice, this threshold is expressed in term of number of $\sigma$.

In this work, DisPerSE is run with a 5$\sigma$ persistence threshold to extract the persistent cosmic web from the density field as computed from the discrete distribution of the galaxies in the SDSS main sample using the Delaunay Tessellation Field Estimator technique (DTFE; \citet{schaap2000}; \citet{cautun2011}). An illustration of the filamentary network overlaid with the density contrast of the underlying galaxy distribution for the SDSS field is shown in Figure \ref{fig:disperse}. 

Being reliant on the three-dimensional distribution of galaxies, DisPerSE is therefore affected by redshift space distortions. On large-scales this corresponds to the Kaiser effect \citep{kaiser1987}, which acts to increase the contrast of the skeleton due the coherent motion of galaxies with the growth of structure \citep[e.g.][]{shi2016}. On small-scales, however, the Fingers of God effect \citep[FOG;][]{jackson1972,tulley1978} derives from random motions of galaxies within virialized haloes. The latter can elongate structure in redshift space leading to erroneous identification of filaments. We correct for the FOG effect using the technique outlined in \citet{kraljic2018}.

\subsection{Cosmic web distances}
Having constructed a skeleton of the cosmic web, a galaxy's environment can be described by finding its vicinity to various features of the skeleton. The cosmic web comprises of low density `void' regions which are enclosed by `walls' of structure which become filaments at points of intersection. The gravitational potential of the filaments dictate the flow of the matter, which at the point of intersection, feed high density regions interpreted as `nodes'. Along the filament, saddle points remain as minima between the flows towards nodes.  

The distance to the nearest filamentary point, $D_{skel}$, is first found for each galaxy. To then consider the influence of the nearest node, the distance from this impact point along the filament to the node is also computed, $D_{node}$. Finally the distance to the nearest wall, $D_{wall}$, can then be found. In order to investigate expected trends of galaxies with vicinity to any cosmic web feature we must remove effects resulting due to the proximity of others. For $D_{skel}$ we remove all galaxies that lie within $D_{node} < 0.5$ Mpc. 
This represents a compromise between eliminating the effect of other cosmic web features and having enough galaxies left to construct a statistically significant sample. Tightening the condition with respect to nodes so that we require $D_{node} > 1$ Mpc does not change any of the results presented in this work. 
We do not include analysis with respect to the walls since we limited to low numbers after removing galaxies that could be influenced by nodes and/or filaments.

Construction of the cosmic web from any observation is influenced by the completeness and the sampling of the galaxy sample. The modified SDSS DR10 spectroscopic sample is complete to $m_r$ = 17.77. A sample containing only brighter galaxies will naturally only identify stronger/larger filamentary features and hence smaller substructures will be missed. In addition, the lower the sampling of galaxies, the lesser the accuracy of the actual position of cosmic web features. To correct for this, the distances are normalised by the mean inter-galaxy separation, $\left\langle D_z \right\rangle$ at a given redshift, as such $\left\langle D_z \right\rangle = n(z)^{-1/3}$ where $n(z)$ is the number density. 

\subsection{Environmental density and stellar mass}
Before we consider the role of the cosmic web, we consider the role of environmental density and stellar mass in our results. A dependence on small scale density is an indirect effect of halo mass and would not probe the large-scale anisotropy of the cosmic web. It is also important to isolate the role of stellar mass and morphology following their distinct gradients with respect to cosmic web features found in GAMA \citep{kraljic2018}.

\begin{figure*}
	\includegraphics[width=\linewidth]{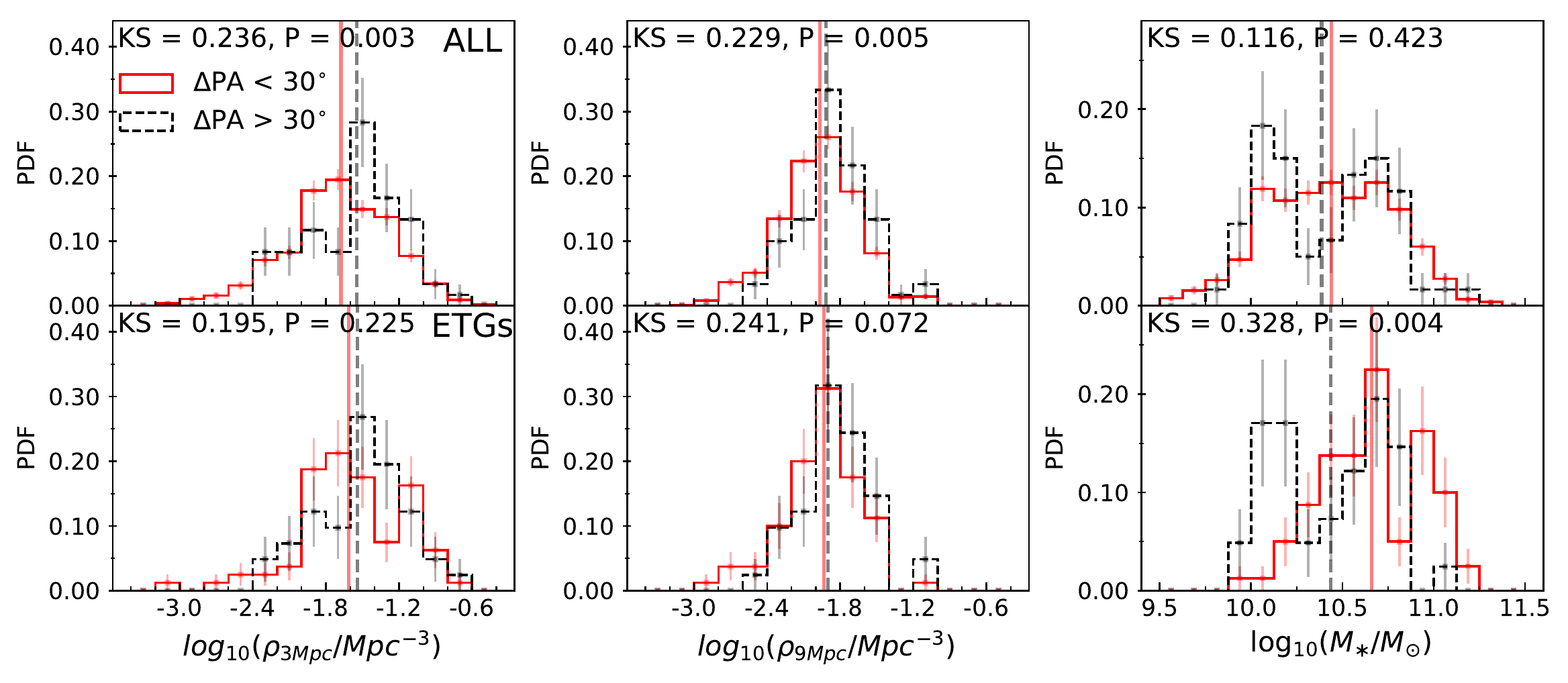}
    \caption{Probability density distributions of density smoothed with a Gaussian kernel at the scale of 3 Mpc \& 9 Mpc and the stellar mass, log$_{10}(M_{\ast}/M_{\odot})$ (left to right) for all $\Delta$PA defined galaxies (top row) and all GalaxyZoo1 classified elliptical galaxies with a $\Delta$PA (bottom row). Aligned galaxies ($\Delta$PA < 30$^{\circ}$) are shown in red (solid line) and those with high misalignment ($\Delta$PA > 30$^{\circ}$) are in black (dashed line). Each histogram is given with Poisson errors on each bin. A two-sample KS statistic and its corresponding p-value is overlaid for comparison between the distributions in each cell and the vertical lines denote the corresponding distribution's median. Using all galaxies, the misaligned sample resides in higher density at the scales of 3 Mpc and 9 Mpc to the aligned sample respectively, but are equivalent in stellar mass. Selecting only ETGs accounts for the difference in small scale density but the misaligned sample are at lower stellar masses than the aligned.}
    \label{fig:density}
\end{figure*}

Figure \ref{fig:density} shows the distributions of densities and stellar mass for the aligned and misaligned samples. Density used in this work is computed using a Delaunay tessellation of the discrete galaxy positions through the DTFE estimator, smoothed with a Gaussian kernel at local scales (3 Mpc) and at large scales (9 Mpc). 

We evaluate the likelihood of the aligned and misaligned sample being drawn from the same continuous distribution through implementation of a two-sample Kolmogorov--Smirnov (KS) test. In each case a D-value of the KS statistic with a corresponding p-value is provided. The D-value (referred to as the KS statistic hence-forth) provides the maximum fractional difference between the cumulative distribution functions with a p-value corresponding to the null hypothesis that the two samples are drawn from the same continuous distribution. A high KS value combined with a low p-value (for example; KS $\geq$ 0.1 for P $\leq$ 10\% confidence level) therefore is consistent with the two samples being significantly different.

The first row of Figure \ref{fig:density} considers the difference between all $\Delta$PA defined galaxies. We find that the misaligned galaxies ($\Delta$PA > 30$^{\circ}$) reside in more dense environments at small and large scales with a probability that the distributions are instead consistent of 0.3\% and 0.5\% respectively. The two samples are consistent in distribution of stellar mass, despite the difference in classified morphology, as shown in Table \ref{tab:GZ}. In the second row of Figure \ref{fig:density} we consider the same properties but only for ETGs. We select on optical morphology using the classifications of GalaxyZoo1 as introduced in \S\ref{sec:samp_sec}. We find that the difference seen in density smoothed on small scales may be explained by morphology as the p-value for the KS test increases, however misaligned galaxies tend to reside in more dense large-scale environments and populate lower stellar masses compared with aligned galaxies of the same morphology.

In order to minimize the effect of $\rho_{3Mpc}$ and $M_{\ast}$ in our cosmic web results, in the next section we weight distance distributions on both stellar mass and small scale density, when comparing the distribution of cosmic distances for the aligned and misaligned samples. This is done through normalising the histogram of the weight quantity to be consistent between distributions using a minimum of three bins. We also include the results of ETGs only to minimize the impact of morphology. We cannot do the same for LTGs due to the lack of misaligned LTGs. 

\subsection{Results of cosmic web distances} \label{sec:cw_res}
\begin{figure}
	\includegraphics[width=\linewidth]{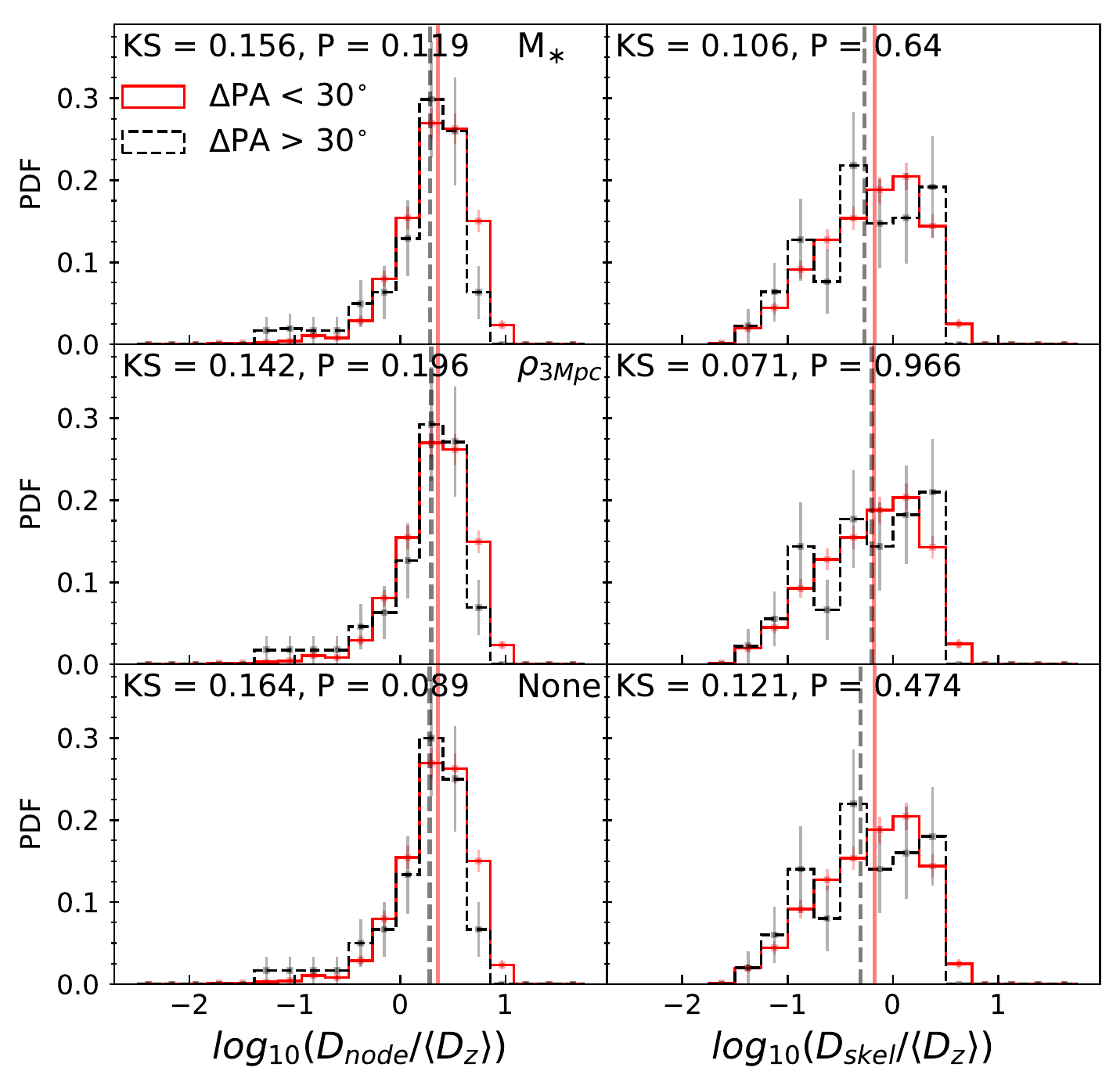}
    \caption{Probability density distributions of normalised distances to cosmic web features for all $\Delta$PA defined galaxies. Distances to nodes (left) and filaments (right) are normalised by the sampling at a given redshift. The distributions of galaxies in the top row is weighted on stellar mass between the aligned (red solid line) and misaligned samples (black dashed line). The distributions are weighted by density smoothed by a Gaussian kernel at the scale of 3 Mpc for the middle row and are left unweighted for the bottom row. A two-sample KS statistic and its corresponding p-value is overlaid for comparison between the distributions in each cell. The error bars represent the poisson noise in each bin. The weighted median values for each distribution are shown by the vertical lines.}
    \label{fig:cw_all}
\end{figure}

Figure \ref{fig:cw_all} shows the distance probability density distributions of aligned and misaligned galaxies with respect to nodes (left) and filaments (right). The top row shows the two samples weighted on stellar mass, the middle row weighted on small scale density (3 Mpc smoothed) and the bottom row shows the raw distributions. The results of a two-sample KS test with corresponding weightings to the cumulative distribution function are overlaid in each cell. 

The distributions of aligned and misaligned galaxies with respect to filaments meet the null hypothesis criterion of high p-values for all weighting schemes (i.e. no statistically significant difference between distributions). This is indicative that $\Delta$PA is independent of the influence of filaments identified in our analysis. For the unweighted samples, we find that misaligned galaxies typically reside in closer vicinity to nodes than their aligned counterparts as indicated by a p-value of 0.089. This difference is however partially negated by weighting on stellar mass or density smoothed on the 3 Mpc scale, as reflected in slightly reduced KS values and p-values increased above the 0.1 significance level.

\begin{figure}
	\includegraphics[width=\linewidth]{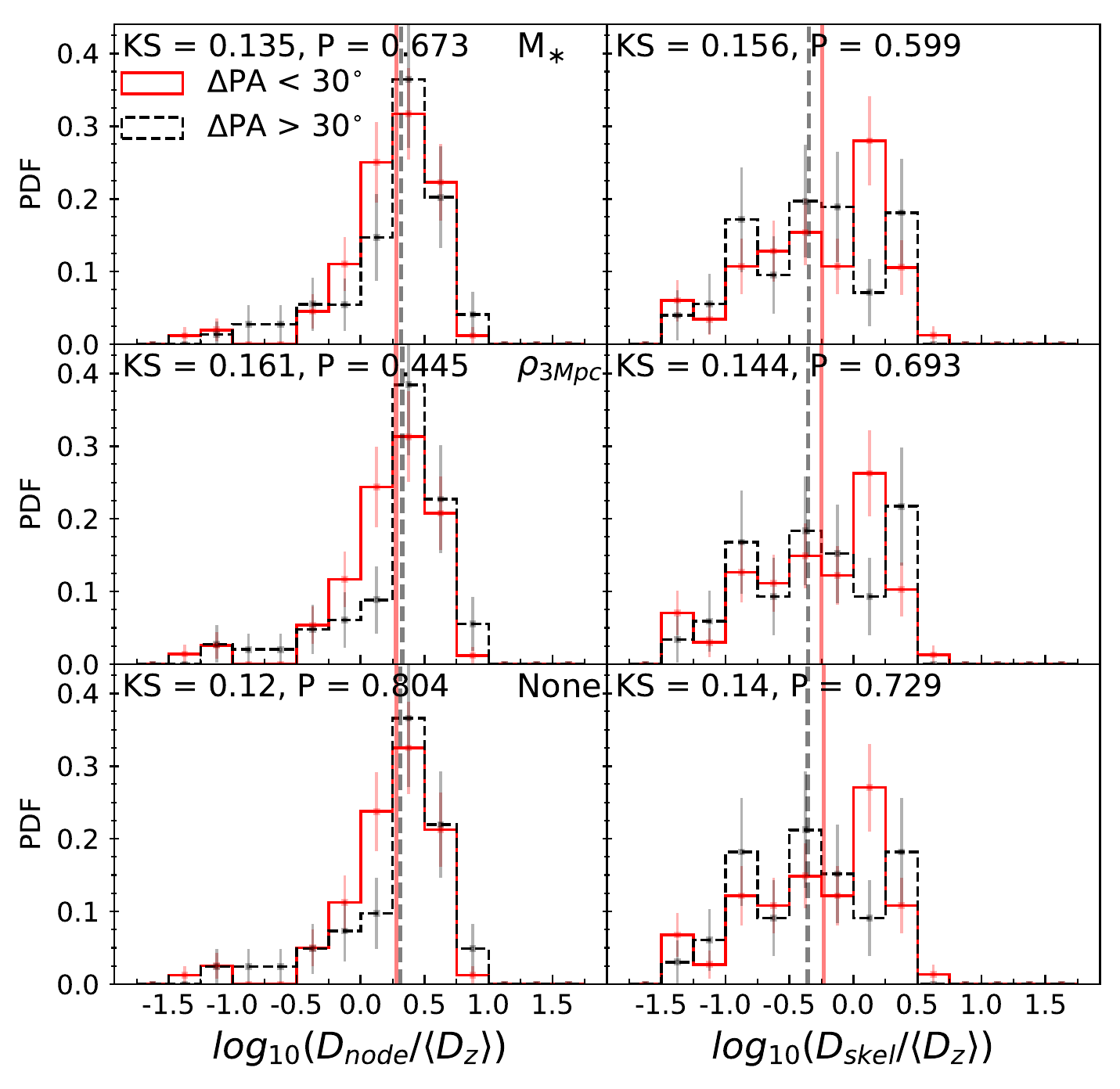}
    \caption{Same as Figure \ref{fig:cw_all} but using only visually selected ETGs as found in GalaxyZoo1.}
    \label{fig:cw_et}
\end{figure}

The origin of misaligned galaxies residing preferentially closer to nodes could be explained by their morphology difference with respect to the aligned sample. In previous work, \citet{kraljic2018} found distinct gradients of stellar mass and morphology with vicinity to nodes and filaments. Figure \ref{fig:cw_et} shows the distributions of cosmic web feature distances but now only selecting ETGs. We find that in all weighting schemes, the distance distributions of aligned and misaligned galaxies with respect to both nodes and filaments meet the null hypothesis criterion as reflected in large p-values (> 0.4). These distributions appear to be drawn from the same continuous distribution, indicating that direct and indirect effects of morphology are likely responsible for the difference in distance to nodes. 

\subsection{The role of halo mass}
Our primary aim in this section is to isolate if vicinity to filaments can impact the rate of present-time accretion on a central galaxy in a low-mass halo. Including high-mass haloes in our sample may counteract any observable signal as they are possible candidates responsible for quenching accretion. High-mass, typically old haloes, are the opposite of what we are trying to target: young, still forming low-mass haloes (with respect to old low-mass haloes). 

\begin{figure}
	\includegraphics[width=\linewidth]{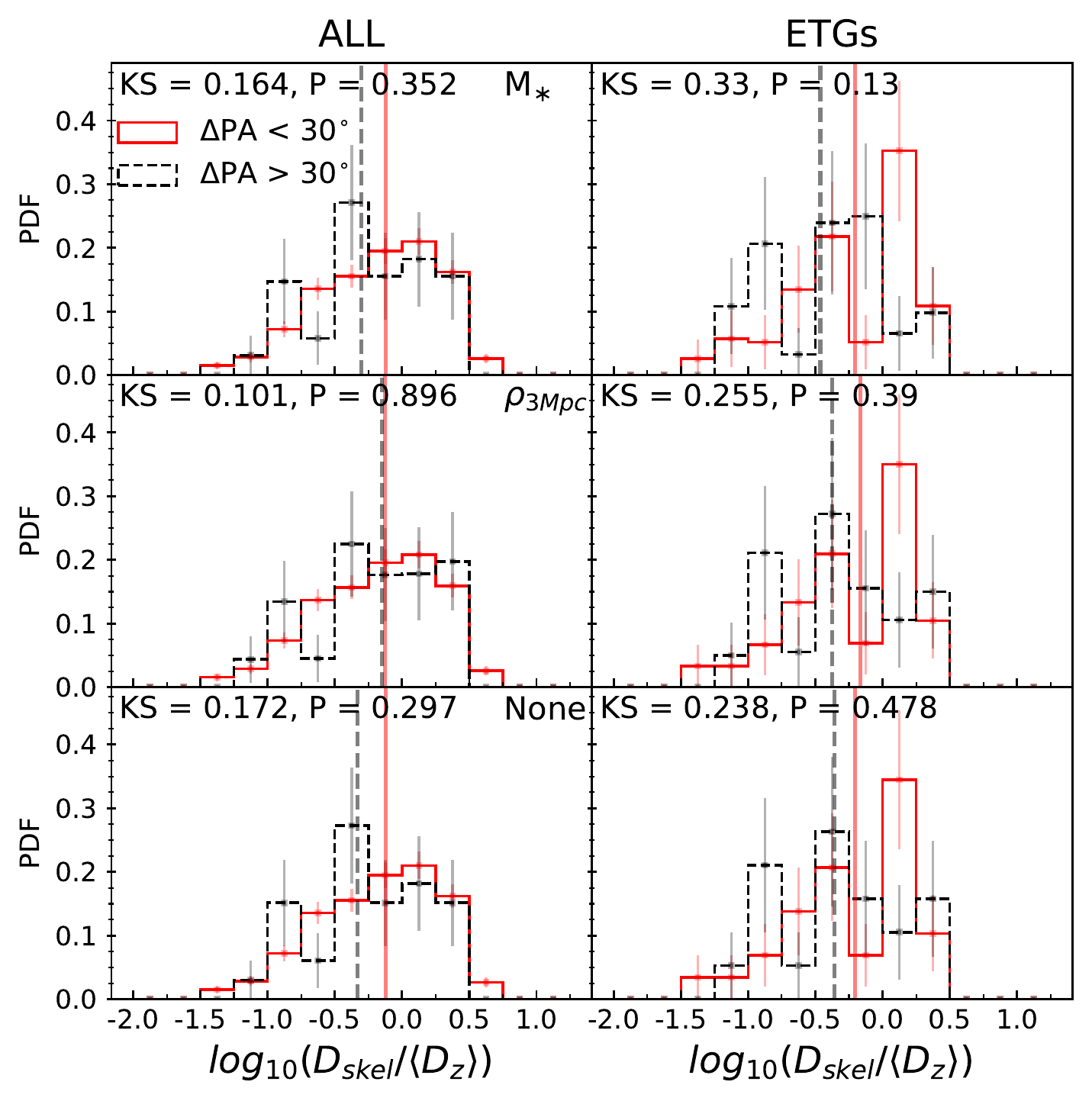} 
    \caption{Probability density distributions of normalised distances to filaments for galaxies with log$_{10}$(M$_h$/M$_{\odot}) \le 12.5$. As in Figure \ref{fig:cw_all} the distributions are weighted on stellar mass (top), density smoothed at scales of 3 Mpc (middle) and left unweighted (bottom). The distributions of all $\Delta$PA defined galaxies (left) and only galaxyZoo selected ETGs (right) are shown for comparison.}
    \label{fig:mh_cw}
\end{figure}

We now consider $D_{skel}$ for low-mass haloes only. \citet{tojeiro2017} found signal of halo assembly bias in low-mass haloes using the stellar to halo mass ratio in GAMA. Low-mass haloes residing in regions of stronger tidal forces were found to form earlier irrespective of density, with this trend apparently reversed at high mass. This signal was found to be strongest for haloes of mass $M_h \sim 10^{12.3} M_{\odot}$, however a slight trend was found even at $M_h \sim 10^{12.74} M_{\odot}$. To ensure we have enough objects for a statistically significant sample we therefore consider all central galaxies residing in haloes of mass: $M_h \sim 10^{12.5} M_{\odot}$. Figure \ref{fig:mh_cw} shows the distance probability density distributions for the whole $\Delta$PA defined sample and GalaxyZoo1 defined ETGs only. For all weighting schemes, we find p-values consistently above the 0.1 significance level and hence conclude the null hypothesis that the aligned and misaligned galaxies are consistent in distance distributions with respect to filaments. This holds true regardless of morphology selection.

\section{Stellar to halo mass ratio}\label{sec:MsMh}
In this section we introduce an observational proxy for halo formation time: the central stellar to total halo mass ratio. Its use was motivated in \citet{wang2011} who explored the correlation between various halo properties which were identified within a set of seven N-body simulations using the P\textsuperscript{3}M code described in \citet{jing2007}. One of the most important properties is its formation time, $z_f$ which was shown to correlate with galaxy properties such as SFR, galaxy age and colour. The formation time in this instance is defined as the redshift at which the main progenitor has formed half of the mass of the final halo. 
They establish that formation time shows a tight correlation with the sub-structure fraction, $f_s = 1 - M_{main}/M_h$, where $M_{main}$ and $M_h$ are the main \textit{sub}-halo mass and the halo mass respectively \citep{gao2007}. 
The ratio of $M_{main}$ and the total halo mass is seen to act as an robust estimate for formation time. Following \citet{lim2015}, we use the following as an observational proxy,
\begin{equation}
f_c = \frac{M_{*,c}}{M_h},
\end{equation}
where $M_{*,c}$ is the stellar mass of the central galaxy. The $M_h$ in this instance is found using the group stellar mass ranking from Y07. $M_{*,c}$ is a reasonable estimator for the main sub-halo mass $M_{main}$ however they do not hold an exactly monotonic relation. Given this and that $f_s$ is not perfectly correlated with formation time, $f_c$ can only be considered to be a relative proxy of $z_f$ as shown in \citet{lim2015}. A higher value of $M_{*,c}/M_h$ should correspond to a relatively older halo. 

Semi-analytic and hydrodynamical simulations have since confirmed a correlation between halo assembly time and the stellar to halo mass ratio, and have shown how halo formation time partly explains the scatter in the stellar mass to halo mass relation \citep[e.g.][]{matthee2017,tojeiro2017,zehavi2018}. Observationally, \cite{tojeiro2017} show that the stellar to halo mass ratio of central galaxies varied with position within the cosmic web, at fixed halo mass. In this section, we investigate whether recent accretion history, associated with younger halos, might be visible in the kinematics of gas and stars.

\subsection{Results of the halo age proxy}
\begin{figure}
	\includegraphics[width=\linewidth]{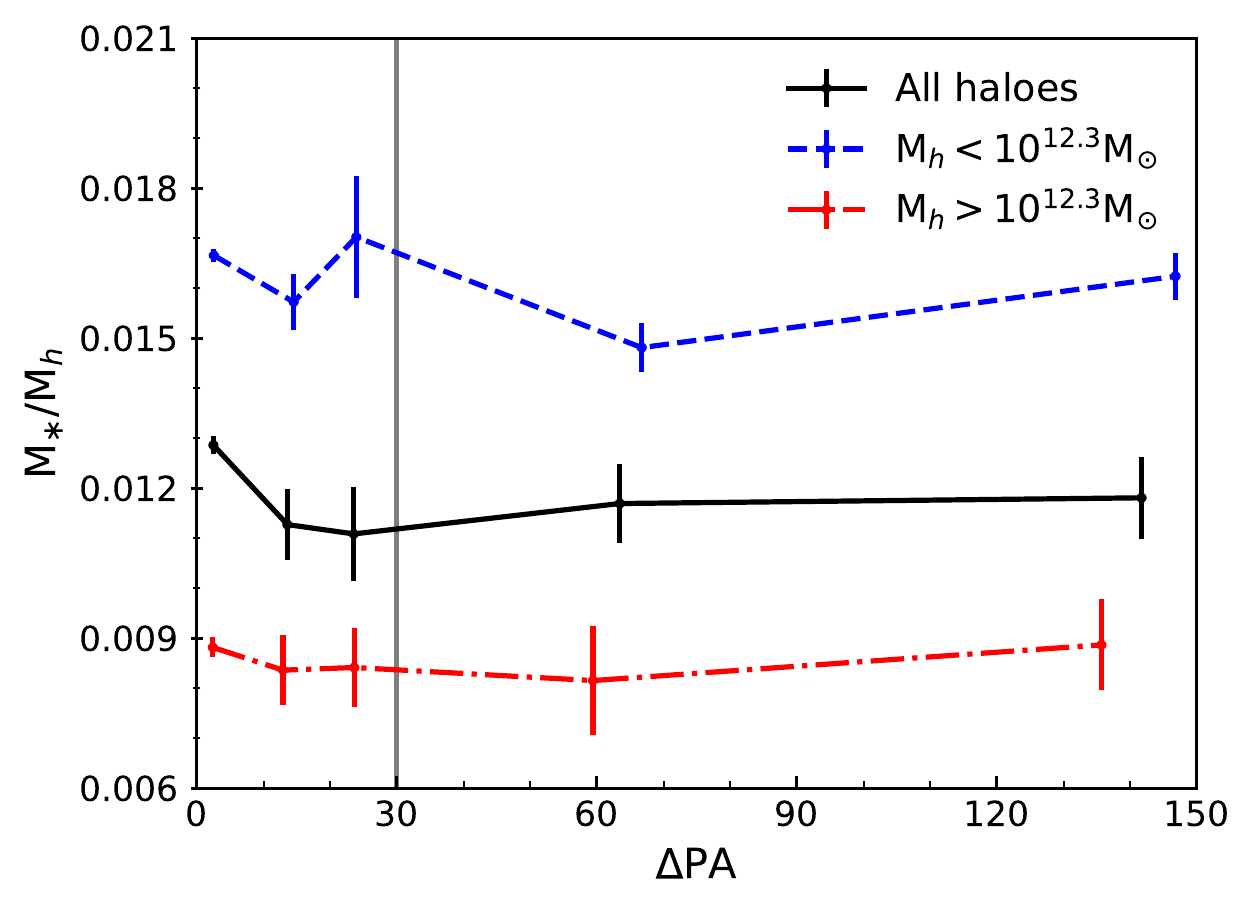}
    \caption{The central stellar to total halo mass ratio for the $\Delta$PA sub-sample in bins of halo mass. Galaxies residing in haloes of M$_{\ast} < 10^{12.3}$M$_{\odot}$ (blue dashed line), M$_{\ast} > 10^{12.3}$M$_{\odot}$ (red dot-dashed line) and the total sample (black solid line) are divided into bins of $\Delta$PA$ = [0,10,20,30,90,180] (^{\circ})$. Error-bars are given by the standard error on the mean. Within each bin of halo mass, distributions are weighted on redshift and halo mass between all $\Delta$PA bins.}
    \label{fig:2d_ratio}
\end{figure}

Figure \ref{fig:2d_ratio} shows the stellar to halo mass ratio as function of $\Delta$PA. Since we do not possess errors for stellar mass or halo mass and can only roughly estimate $\Delta$PA errors, we bin our data and calculate the standard error on the mean. We split our sample at the median halo mass of $M_{\ast} = 10^{12.3} M_{\odot}$ and divide galaxies into bins with boundaries; $\Delta$PA$ = [0,5,10,20,90,180] (^{\circ})$. In each of the halo mass bins, we weigh the redshift and halo mass distributions to be consistent in each bin of $\Delta$PA. Figure \ref{fig:2d_ratio} shows no particular dependence on $\Delta$PA for this proxy of halo age. However, given the strong dependence of $M_{\ast}/M_{h}$ on $M_{h}$, we investigate this further by simultaneously considering the relationship between $M_{h}$, $M_{\ast}/M_{h}$ and $\Delta$PA. 

We divide our parameter space into quarters by splitting galaxies at $\Delta$PA = 30$^{\circ}$ and $M_{h} = 10^{12.3}M_{\odot}$. In each region we fit a flat plane to our data points as described by,
\begin{equation}
M_{\ast}/M_{h} = c_{0} log_{10}(M_{h}/M_{\odot}) + c_{1}\Delta PA + c_{2}.
\end{equation}
A strong correlation between $M_{\ast}/M_{h}$ and $\Delta$PA would correspond to a relatively large value of $c_{1}$ with regards to $c_{0}$. To understand the significance of any result, we also fit a flat plane with $c_{1} = 0$ (i.e. no dependence on $\Delta$PA) and evaluate $\chi_{red}^2$ for both. These values are found in Table \ref{tab:chisq}. We are inherently limited by having no errors on the estimates of stellar mass and halo mass for our sample. We therefore construct constant errors across the sample estimated from the sample variance. The sample variance itself is found from considering each data point with regards to its 10 nearest neighbours in the parameter space. 

We find that fixing the gradient along $\Delta$PA has little or no effect on the fit of the linear plane, regardless of how we sub-divide our parameter space. In some cases the comparison planes are effectively the same allowing for a smaller $\chi_{red}^2$ for the two free parameter fit. As discussed in \S\ref{mass}, halo masses assigned to galaxy groups using the Y07 group catalogue are corrected due to incompleteness above $z=0.09$. We consider the plane fitting again but with redshift cuts at both $z=0.09$ and a conservative $z=0.05$. In both instances, we also find there are no statistically significant gradients along $\Delta$PA. We therefore conclude that $\Delta$PA holds little correlation with the age of the halo in which it resides, as inferred from current measurements of $M_{\ast}/M_{h}$. 


\begin{table}
\begin{tabular}{|l|c|c|c|}
\hline
$M_{h}/M_{\odot}$& & $\Delta$PA = [0, 30]($^{\circ}$) & $\Delta$PA = [30, 180]($^{\circ}$) \\ \hline 
[$10^{11.7}, 10^{12.3}$] & $c_{1} = 0$: & 1.188 & 1.070 \\
					   & Free : & 1.190 & 1.068 \\ \hline
[$10^{12.3}, 10^{14}$]   & $c_{1} = 0$: & 1.218 & 0.988 \\ 
					   & Free: & 1.212 & 1.001 \\ \hline
\end{tabular}
\caption{$\chi_{red}^2$ for plane fits with $c_1 = 0$ and left free in the parameter space for the central stellar to halo mass ratio (M$_{\ast}$/M$_{h}$), halo mass (M$_{h}$) and $\Delta$PA. The parameter space is divided at $\Delta$PA = 30$^{\circ}$ and $M_{h} = 10^{12.3}M_{\odot}$.}
\label{tab:chisq}
\end{table}

\section{Halo occupation distribution}
In this section we introduce the HOD function and how it can be used to infer halo age. In describing the relationship between galaxies and dark matter haloes, HODs are a useful prescription to determine models of galaxy formation and evolution \citep[e.g.][]{berlind2003}. They provide a probability distribution function $P(N|M_h)$ for a set of virialised haloes where $N$ is the number of hosted galaxies for a given halo mass $M_h$. A fundamental assumption underlying HOD modelling is that the galaxy occupation is purely dependent on the halo mass. Typically the observed galaxy clustering is used to construct the empirical relationship that allows mock dark matter haloes to be populated with galaxies. Assembly bias would directly affect the observed clustering of galaxies and hence challenge any interpretation using the HOD framework. 

Continuing our discussion, low-mass haloes near large haloes are expected to cease formation earlier. This leads to a boost of galaxy clustering at this halo mass range relative to the overall sample as they live preferentially in high density regions. \citet{zehavi2018} previously investigated the dependence of occupation functions on various properties such as large-scale environmental density and halo age using semi-analytical galaxy models applied to the Millennium simulation \citep{springel2005} \citep[See also;][who confirmed these results using the hydro simulations of EAGLE and Illustris]{artale2018}. They find that higher density environments generally act to populate lower mass haloes with central galaxies. A stronger dependence can be found on halo age, however, as earlier forming low-mass haloes are more likely to host central galaxies. In addition, earlier forming haloes are likely to host fewer satellites relative to late forming haloes at fixed halo mass. A simple explanation is that the early forming haloes provide more time for their constituent satellite galaxies to merge with the central. More massive central galaxies may therefore reside in low-mass haloes that formed early due to this general in-flow, analogous to a higher stellar to halo mass ratio. 

\subsection{Background subtraction}
To understand the assembly history of a central galaxy's sub-halo we must consider the role of satellites that contribute to the hierarchical structure growth of its halo merger tree. However, we are limited by the magnitude and typical size of galaxies inhibiting small substructure around a main sub-halo. \citet{liu2011} demonstrate a common method for counter-acting the lack of spectroscopic information for satellite galaxies through counting possible photometric group members. Their numbers are then statistically corrected to remove the contribution of contaminant foreground and background galaxies outside of the group. This enables a lower limit of apparent magnitudes which can be accessed through use of the background subtraction technique. \citet{rodriguez2015} extend this formalism to HOD modelling and provide the technique we implement here. For a complete description of the technique we direct the reader to this reference, however we will summarise the basic concepts here.

Background subtraction requires two catalogues that share the same sky area; we will use our $\Delta$PA defined MaNGA centrals with their identified groups in combination with the photometric SDSS catalogue. For the photometric galaxies in the sky region of a group, their absolute magnitudes are calculated at the redshift of the group, $z_{f}$. The total number of galaxies with an absolute magnitude $M \le M_{min}$ are then counted within a circle around the group centre with its radius determined by the projected characteristic radius on the sky. In order to remove background galaxies, an estimation of the local density with respect to the average catalogue density must be made. All galaxies with $M \le M_{min}$ are recounted in concentric annuli centred on the group to provide the local density. A correction for the total number of galaxies in the group can then be estimated by subtracting the local background density multiplied by the group's projected area. The HOD is then constructed by binning the groups into mass intervals and averaging.

\citet{rodriguez2015} demonstrate the recovery of the background subtraction method using mock catalogues constructed from semi-analytic models of galaxy formation applied on top of the Millennium simulation. They compare the HODs found from the background subtraction technique to HODs of the direct galaxy counts in volume limited samples for different magnitude limits \citep[see Figure 1 in;][]{rodriguez2015}. Beyond a small overprediction for fainter magnitudes ($M_{lim} \approx -16.0$), they find good agreement with direct galaxy counts for all absolute magnitudes.

Results from the background subtraction technique have also been compared to results of other HOD estimation techniques applied to observations. \citet{yang2008} parametrise HODs for satellite galaxies in groups identified in SDSS DR4 using the adaptive group-finding algorithm of Y07 (see \S\ref{mass} for discussion). The background subtraction technique shows great agreement in estimating parameters of the HOD relative to the method of \citet{yang2008} but additionally offers the ability to estimate the HOD for fainter absolute magnitudes than previous work \citep[see Figure 5 in][]{rodriguez2015}.

MPL-6 does not provide a large enough sample size to construct a reliable two-halo term in HOD modelling through calculation of the cross-correlation function. Background subtraction therefore represents the best estimation for this sample size.

\subsection{Results of the HOD}
We match each central galaxy in our $\Delta$PA sample with its corresponding satellite group members using \citet{yang2007}. We split our groups at $\Delta$PA = 10$^{\circ}$ for the central and calculate the HOD using the background subtraction technique. Our lower split in $\Delta$PA is purely due to limitations of sample numbers. As demonstrated in \S \ref{def_mis}, this should be above the resolution limit of $\Delta$PA, however may include more galaxies with spurious kinematic misalignments or due to an internal origin. 

\begin{figure*}
	\includegraphics[width=\linewidth]{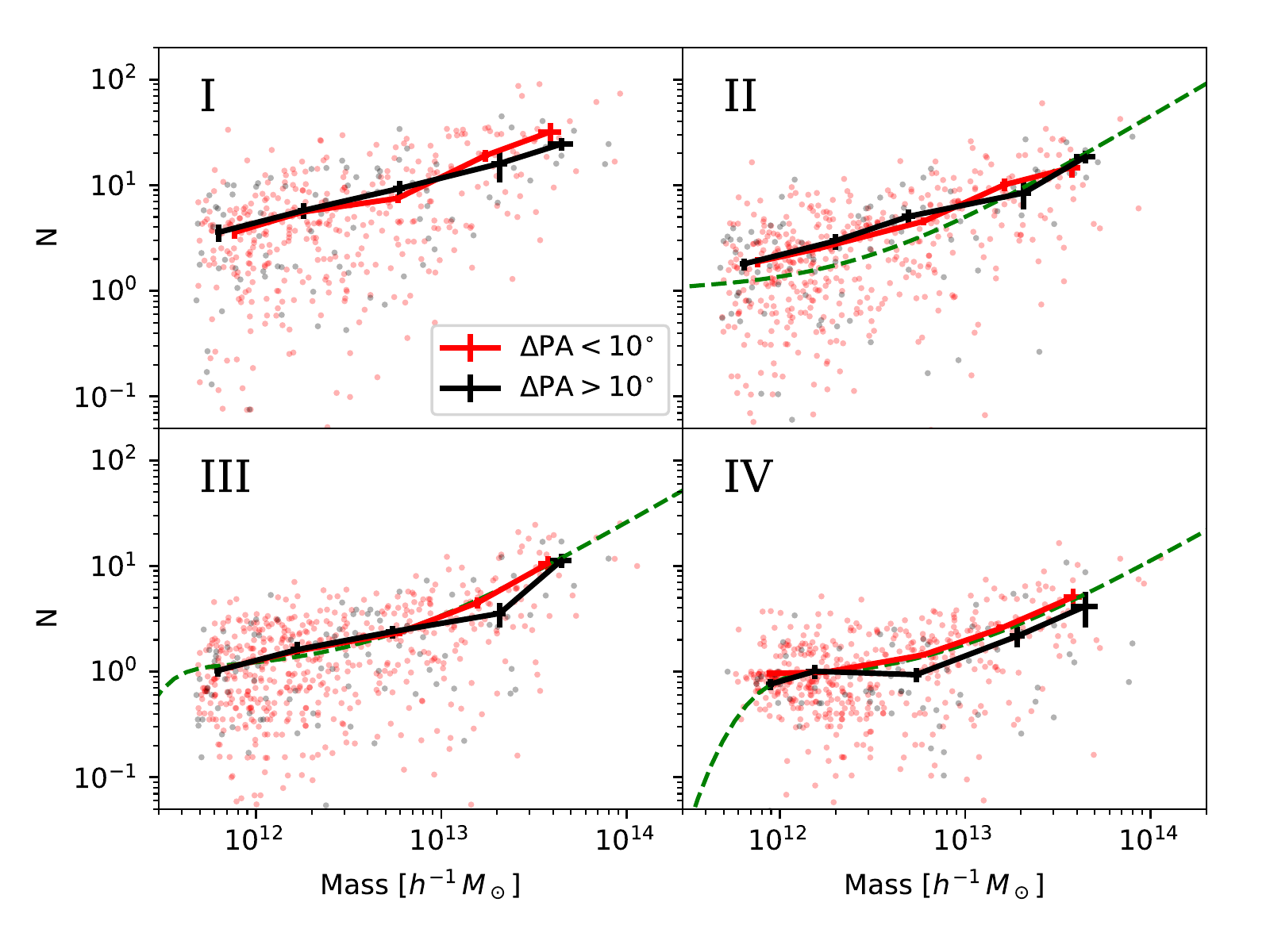}
    \caption{Halo occupation distributions using background subtraction for groups with central galaxies with $\Delta$PA < 10$^{\circ}$ (red) and $\Delta$PA > 10$^{\circ}$ (black). Panels I, II, III \& IV correspond to a $r$-band magnitude cut of $\leq$ -17, -18, -19 and -20 respectively. The points reflect the estimation for individual groups, with these lines representing the mean with corresponding errors on the mean. \blue{In panels II-IV, comparison to HODs estimated directly from clustering with similar magnitude cuts are shown by the green dashed lines \citep[][see text]{zehavi2011}.}}
    \label{fig:hod_mpl6}
\end{figure*}

Figure \ref{fig:hod_mpl6} shows the HOD for different magnitude cuts of the group during comparison to the photometric background. As fainter galaxies are removed, the overall magnitude of the HOD naturally decreases as we have less complete groups. 
\blue{As a sanity check, we compare our HODs estimated from the background subtraction technique to HODs estimated directly from clustering in SDSS DR7 with similar magnitude cuts \citep{zehavi2011}. The authors use measurements of the projected correlation function for SDSS DR7, which is translated into a HOD through use of a smoothed step function \citep[see equation 7;][]{zehavi2011}. This comparison is shown in panels II-IV (green dashed line) and matches our estimation well.}
Regardless of $\Delta$PA classification, at all magnitude thresholds we find that the difference between the HODs are indiscernible. We conclude that $\Delta$PA of the central galaxy does not produce a difference in its constituent group assembly that can be seen through occupation functions. A detailed analysis using cross-correlations will be presented in a future paper, once the sample size of MaNGA is sufficient.

\section{Discussion}
An important assumption in our motivation has been that gas accretion should originate from cold filamentary flows of the cosmic web. In reality, gas in-flowing into a central galaxy could also result from cooling of the surrounding hot halo and accreting in a more stochastic nature, so we consider that next. 

As introduced in \S \ref{def_mis}, \citet{lagos2015} explore the origin of kinematic misalignment between gas and stars in ETGs using GALFORM, in comparison with the misaligned field ETG fraction found in ATLAS\textsuperscript{3D} of 42 $\pm$ 6\%  \citep{davis2011a}. They find that using solely galaxy mergers as the source for misaligned cold gas only predicts 2\% of ETGs to have $\Delta$PA > 30$^{\circ}$. Regardless of the time-scales of dynamical friction used, there are simply not enough mergers at $z=0$ to recreate the misaligned fraction observed. To include the effects of smooth gas accretion, they trace its history onto a subhalo along with incident galaxy mergers in their GALFORM model. They follow the angular momentum flips in the constituent cold gas, stars in galaxies and the corresponding dark matter halo using the Monte Carlo simulation prescription of \citet{padilla2014}. This simulation analyses the incident mass with respect to the subhalo, categorising by source (smooth accretion or merger) and constructs a PDF of the expected change in rotational direction for each component: hot halo, cold gas disc and stellar disc. \citet{padilla2014,lagos2015} consider the gas and stellar discs of galaxies to be initially aligned with the surrounding hot halo of gas from which they cooled. When a dark matter halo is accreted, the hot halo is immediately offset from the original rotation, which in time cools to create a misaligned gas disc in the galaxy. Memory of the misalignment can be erased through disc instabilities which use cold gas in the form of a starburst. It should however be noted that this model does not include the relaxation of the gas disc towards the stellar component due to torques. With this in mind any expected misalignment can only be considered an upper limit. \citet{lagos2015} reproduce consistent fractions of misalignment with ATLAS\textsuperscript{3D} by assuming that accretion does not come from a correlated preferential direction. To consider the effect of filamentary `cold mode' accretion on misalignment, the direction of accretion is then correlated on various time-scales and again the expected misaligned fraction is calculated. Assuming an uncorrelated direction of accretion marginally better reproduces observations but more importantly highlights the important role of slower `hot mode' accretion in interpretation. Stochastic accretion onto the galaxy from the hot halo may be the driving factor in misalignment of the gas disc, explaining the lack of correlation with our measures of large-scale environment and halo age. 

\citet{correa2018} investigated the role of cold and hot modes of accretion onto galaxies with respect to the accretion rate onto the host DM halo using the EAGLE suite of hydrodynamical cosmological simulations. In haloes of mass > $10^{12}M_{\odot}$, the two modes of accretion coexist and both contribute to the gas accretion rate on central galaxies. Below this value the cold mode of filamentary flows appear to dominate whereas the hot mode dominates above $10^{12.7}M_{\odot}$ for $z = 0$.  They note that AGN feedback plays an important role on the ability of gas from the surrounding hot halo to cool and accrete and is likely less efficient at high halo masses explaining why hot mode accretion becomes dominant. The ability of cold flows to reach the halo centre is, however, unconfirmed. \citet{nelson2013} find that the majority of gas from cold mode accretion is shock heated as it travels from the DM halo. They compare the differences of the moving mesh code AREPO with the results of GADGET-3 using otherwise identical simulation runs. While gas filaments in GADGET remain collimated and flow coherently to small radii, the same filamentary gas streams in AREPO are heated and become disrupted around  0.25-0.5 r$_{vir}$, boosting the rate of hot gas accretion as a result. The prominence of cold and hot modes of accretion and their subsequent ability to misalign the gas component of a galaxy is rightfully under debate. The lack of correlation of kinematically misaligned galaxies with environments of expected continued accretion could simply indicate that hot mode accretion is dominant in these regimes. 

Another consideration is the visibility of accretion within the effective radii observed in MaNGA. The Primary+ galaxy sample (63\% of MaNGA total including both Primary and Colour Enhanced samples) observes galaxies up to a minimum of 1.5 $R_e$, whereas the Secondary sample (37\% of MaNGA total) goes to a minimum of 2.5 $R_e$. All modes of accretion would be expected to be most visible on the outskirts of the central galaxy which could be further than 1.5-2.5 $R_e$. Below this we could expect gas and stellar components to align on much faster time-scales after an accretion event due to the strength of stellar torques peaking closer to the galactic centre. Despite this, it should be noted that approximately only 20\% of galaxies with $\Delta$PA > 30$^{\circ}$ are from the Secondary sample whereas the fraction of aligned galaxies in the Secondary is as expected from the targeting. 

To assess the impact of changing the observation extent between 1.5 and 2.5 $R_e$ on $\Delta$PA, we consider all Secondary sample galaxies. We find no significant difference in $\Delta$PA when fitting to an aperture 0.6 (1.5/2.5) of the total size of the original IFU extent. This could be a natural limitation of $\Delta$PA being an average property over all radii and hence being preferentially biased towards the rotation of its likely kinematically aligned centre when considering a population excluding recent mergers. The probability of misalignment is linked to the mass of accreted material and lower mass accretion may well propagate to `warps' in the gas velocity map (i.e. $\Delta$PA changing as a function of radius) while maintaining an aligned classification. During visual inspection we found the scenario of a warped gas map while maintaining an undisturbed stellar velocity field to be rare (seen in approximately 20 galaxies). Bars could also be attributed to create warps in velocity fields. We look to \citet{stark2018} who implement a modified radon transform to characterise PA and its radial variation in the velocity fields of MaNGA for the prevalence of these effects.

Finally, we consider the impact of using a group catalogue to identify central galaxies and estimate halo masses. As discussed in \S\ref{mass}, halo mass is less accurate for small groups, and central mis-classification is more problematic at large halo mass. An estimate of halo mass is only important in one of our tests, where we consider the stellar to halo mass ratio as a proxy for halo formation time. It is possible that errors in halo mass estimates simply averaged out any real signal of $\Delta$PA with the stellar to halo mass ratio. Whereas our two other tests use halo mass estimates to split the data into two populations, the dependence on halo mass values is much reduced, and the Y07 catalogue has been shown to reproduce general trends of galaxy properties as a function of halo mass well \citep{campbell2015}. 
Mis-classification of central galaxies has implications throughout our paper. However at $M_h < 10^{13} M_{\odot}$, where effects of halo assembly are expected to be more prominent, the fraction of groups where stellar ranking results in mis-identification of a satellite as a central is estimated to be well within 10\% by \cite{campbell2015} and \cite{reddick2013}. \blue{To consider how 10\% mis-classifications could impact Figure \ref{fig:2d_ratio}, we perform 50 realisations where we remove 10\% of our central sample and replace these with satellites (with their own defined $\Delta$PA) with a consistent distribution in halo mass. The overall amplitude of $M_{\ast}/M_{h}$ tends to decrease (especially for low halo mass groups), however there appears to be no noticeable changes in trend. This is expected if a signal is not strong with either population of galaxies.}

Although a quantitative assessment of the effects of the group catalogue can only be made using a forward-model approach using mock catalogues we argue, based on the above, that the lack of signal reported in this paper is more likely due to a lack of physical correlation between halo assembly history and kinematic misalignment measured up to 2.5 $R_e$. 

\section{Conclusions}
In this paper, we considered the visibility of cosmic web accretion and hence halo assembly onto central galaxies in MaNGA. We used the difference in global position angles measured for the stellar and H$\alpha$ velocity fields to classify if a galaxy is kinematically misaligned ($\Delta$PA > 30$^{\circ}$). Our paper is summarised as follows:
\begin{itemize}
\item We first correlated distances to cosmic web features such as nodes and filaments to the aligned and misaligned galaxy samples. We considered the theory that low-mass haloes embedded in filaments (or in close vicinity) find their accretion `stalled' as material moves preferentially towards larger sub-haloes along the filament. This would correspond to aligned central galaxies in low-mass haloes residing closer to filamentary structures. We find that kinematic misalignment holds little or no correlation with the vicinity to nodes or filaments once the effects of morphology, stellar mass and small scale density are considered, as shown in Figure \ref{fig:mh_cw}. 
\item We secondly correlated a proxy for halo age; the central stellar mass to total halo mass ratio, with kinematic misalignment. We explored the idea that large-scale tidal forces dictate the formation time-scales of low-mass haloes ($\lesssim$ 10$^{12.3} M_{\odot}$) which should be reflected both in the halo age but also the likelihood of on-going filamentary accretion being quenched. We found that the magnitude of kinematic misalignment held little or no relation to the proxy of the halo age, as shown in Table \ref{tab:chisq}. 
\item We finally considered the halo occupation distribution as a measure of halo age with older haloes providing more time for satellites to merge and hence decrease the magnitude of the HOD \citep[e.g.][]{zehavi2018}. We estimate the HODs using the background subtraction technique for the aligned and misaligned groups with application of stellar mass weightings between the samples. Regardless of the magnitude limit imposed, we find no statistically significant difference between the groups containing aligned and misaligned galaxies, as seen in Figure \ref{fig:hod_mpl6}. We note in this analysis we split at $\Delta$PA = 10$^{\circ}$ in order to construct a sample size large enough for comparison. While this difference is likely well above the expected average error in $\Delta$PA, internal processes may be erroneously included.
\end{itemize}

We note that the lack of correlation could be indicative that the role of `hot mode' accretion from the cooling of the hot halo may play a far larger role than `cold mode' accretion deriving from the cosmic web flows, even at lower halo masses. The ability of integral field spectroscopy to resolve positions of properties such as gas-phase metallicity and star formation rate histories with respect to the surrounding large-scale environment should shed light on the exact origin of misalignment in future MaNGA studies.

It is important to consider the role of morphology in the interpretation of our results. We find that practically none of our kinematically misaligned galaxies are classified to be LTGs by the visual inspection of GalaxyZoo1. In our future work, we plan to investigate the intrinsic relationship between angular momentum and the likelihood of kinematic misalignment. We will test if this correlation can be reproduced in hydrodynamical simulations and if this distinction is seen in other morphological measures. 

\begin{table*}
\begin{tabular}{|l|c|c|c|c|c|c|}
\hline
Plate-IFU & MaNGA-ID & $\Delta$PA ($^{\circ}$)& RA ($^{\circ}$)& DEC ($^{\circ}$)& $z$ & $n$ \\ \hline
8442-12701 & 1-409576 & 3.0 & 197.668780 & 32.228180 & 0.048906 & 1.138390 \\
8085-12701 & 1-38368 & 3.0 & 51.409941 & -1.030121 & 0.028917 & 0.975738 \\
9888-12701 & 1-593975 & 5.0 & 235.475823 & 28.133979 & 0.033220 & 2.828310 \\
9869-12702 & 1-211257 & 3.0 & 247.241677 & 39.319088 & 0.033924 & 2.001890 \\
9870-12702 & 1-198938 & 20.1 & 230.702023 & 43.773505 & 0.039355 & 2.591970 \\
9488-12702 & 1-384128 & 0.0 & 126.610962 & 20.798135 & 0.025302 & 1.324370 \\
9492-12702 & 1-297860 & 3.0 & 117.422630 & 18.861800 & 0.115710 & 6.000000 \\
9864-12703 & 1-245736 & 5.0 & 216.197425 & 52.088435 & 0.045123 & 0.595875 \\
9024-12704 & 1-314449 & 2.0 & 222.961188 & 33.182981 & 0.102298 & 2.089170 \\
9883-12704 & 1-176925 & 0.0 & 256.550767 & 33.555230 & 0.081944 & 2.697350 \\
7968-12704 & 1-180621 & 13.0 & 324.259881 & 0.428240 & 0.050925 & 4.115960 \\
9507-12705 & 1-585206 & 8.0 & 129.520694 & 25.329505 & 0.028176 & 2.029030 \\
9195-12705 & 1-602993 & 3.0 & 28.748306 & 13.482961 & 0.021289 & 1.361890 \\
9181-12705 & 1-548626 & 2.0 & 120.557871 & 37.150076 & 0.083836 & 1.680350 \\
9000-12705 & 1-149896 & 0.0 & 173.194517 & 52.940898 & 0.027494 & 0.838749 \\
9497-1901 & 1-217527 & 27.1 & 117.375207 & 21.752790 & 0.023729 & 1.075320 \\
9042-1902 & 34-28 & 2.0 & 233.422806 & 28.145483 & N/A & N/A \\
9508-1902 & 1-298864 & 3.0 & 126.198239 & 25.225081 & 0.043775 & 5.986280 \\
8932-3701 & 1-456306 & 2.0 & 194.655340 & 27.176555 & 0.025568 & 1.015260 \\
8443-3701 & 1-422976 & 16.0 & 207.074292 & 25.123889 & 0.029748 & 4.033440 \\
9490-3701 & 1-383608 & 3.0 & 122.352258 & 19.628197 & 0.044676 & 3.136980 \\
9500-3701 & 1-385637 & 7.0 & 132.039142 & 24.842770 & 0.086110 & 2.125210 \\
8944-3701 & 1-390232 & 3.0 & 148.484494 & 34.315304 & 0.039737 & 1.260910 \\
9888-3701 & 1-593989 & 0.0 & 236.008028 & 27.699311 & 0.032192 & 3.674540 \\
8154-3701 & 1-37126 & 3.0 & 44.735649 & -0.766114 & 0.043508 & 2.756150 \\
8322-3702 & 1-575209 & 40.2 & 200.120280 & 31.351993 & 0.046406 & 5.339360 \\
9195-3702 & 1-42250 & 26.1 & 27.842784 & 13.060335 & 0.064157 & 3.214610 \\
8443-3702 & 1-423101 & 11.0 & 207.067074 & 25.733487 & 0.036999 & 6.000000 \\
8153-3702 & 1-36779 & 5.0 & 39.461392 & 0.574641 & 0.062996 & 6.000000 \\
8993-3703 & 1-173384 & 1.0 & 164.048899 & 46.880667 & 0.028590 & 3.275440 \\
9491-3704 & 1-383107 & 22.1 & 120.565822 & 18.402295 & 0.039008 & 2.700600 \\
8444-3704 & 1-421226 & 0.0 & 202.459180 & 31.421014 & 0.024702 & 1.163740 \\
9044-6101 & 1-316023 & 1.0 & 230.686984 & 29.769601 & 0.022916 & 1.774900 \\
8443-6101 & 1-592128 & 150.8 & 207.628850 & 24.976623 & 0.029674 & 6.000000 \\
9491-6101 & 1-382712 & 4.0 & 119.174377 & 17.991168 & 0.041245 & 1.067570 \\
9043-6102 & 1-632643 & 2.0 & 231.304721 & 28.359465 & 0.063601 & 5.432590 \\
9870-6103 & 1-569138 & 0.0 & 233.228291 & 44.538732 & 0.037122 & 4.402430 \\
9871-6104 & 1-322262 & 87.4 & 228.984597 & 43.166780 & 0.017980 & 6.000000 \\
9088-6104 & 1-296457 & 5.0 & 243.833726 & 26.624540 & 0.032197 & 4.627260 \\
8990-6104 & 1-174914 & 1.0 & 174.745256 & 50.005889 & 0.046640 & 1.026420 \\
8311-6104 & 1-568584 & 0.0 & 205.282731 & 23.282055 & 0.026353 & 1.871530 \\
8940-6104 & 1-584801 & 16.0 & 122.433990 & 25.881607 & 0.025651 & 3.367540 \\
9047-6104 & 1-270129 & 6.0 & 248.140879 & 26.380740 & 0.058614 & 1.921970 \\
9024-9101 & 1-262322 & 0.0 & 224.024527 & 34.481831 & 0.065684 & 2.460460 \\
8983-9101 & 1-568624 & 3.0 & 205.129395 & 26.307357 & 0.062564 & 3.671430 \\
8993-9102 & 1-277257 & 1.0 & 165.910107 & 45.179967 & 0.020498 & 6.000000 \\
8139-9102 & 1-71108 & 2.0 & 114.756034 & 31.914709 & 0.040488 & 4.716000 \\
9508-9102 & 1-298842 & 2.0 & 126.278363 & 25.900785 & 0.086262 & 2.326020 \\
\end{tabular}
\caption{Table containing all $\Delta$PA defined galaxies which were classified as undergoing mergers or had a high probability of interaction and hence were removed from our sample as described in \S\ref{sec:samp_sec}. Column (1) is the Plate-IFU identifying the unique observation in MaNGA, column (2) is the MaNGA-ID referring to a unique target within MaNGA, column (3) is $\Delta$PA as given in equation \ref{eq:delPA}, column (4) is the right ascension in degrees, column (5) is declination in degrees, column (6) is the NSA redshift and column (7) is the NSA S\'ersic index.}
\label{tab:interact}
\end{table*}

\section*{Acknowledgements}
CD acknowledges support from the Science and Technology Funding Council (STFC) via an PhD studentship (grant number ST/N504427/1). RT acknowledges support from the Science and Technology Facilities Council via an Ernest Rutherford Fellowship (grant number ST/K004719/1). MAS acknowledges the fellowship grant `ESTADIA BREVE EN EL EXTERIOR' given by CONICET. VW acknowledges support of the European Research Council via the award of a starting grant (SEDMorph; P.I. V. Wild). IL acknowledges partial financial support from PROYECTO FONDECYT REGULAR 1150345. We also thank the anonymous referee for the very helpful comments which improved the presentation of this paper. CD thanks M. Graham for use of a code snippet to overlay hexagonal IFU footprints on Figure \ref{fig:cutout_wIFU}. Funding for the Sloan Digital Sky Survey IV has been provided by the Alfred P. Sloan Foundation, the U.S. Department of Energy Office of Science, and the Participating Institutions. SDSS-IV acknowledges support and resources from the Center for High-Performance Computing at the University of Utah. The SDSS website is www.sdss.org. 

SDSS-IV is managed by the Astrophysical Research Consortium for the Participating Institutions of the SDSS Collaboration including the Brazilian Participation Group, the Carnegie Institution for Science, Carnegie Mellon University, the Chilean Participation Group, the French Participation Group, Harvard-Smithsonian Center for Astrophysics, Instituto de Astrof\'isica de Canarias, The Johns Hopkins University, Kavli Institute for the Physics and Mathematics of the Universe (IPMU) / University of Tokyo, Lawrence Berkeley National Laboratory, Leibniz Institut f\"ur Astrophysik Potsdam (AIP), Max-Planck-Institut f\"ur Astronomie (MPIA Heidelberg), Max-Planck-Institut f\"ur Astrophysik (MPA Garching), Max-Planck-Institut f\"ur Extraterrestrische Physik (MPE), National Astronomical Observatory of China, New Mexico State University, New York University, University of Notre Dame, Observat\'orio Nacional / MCTI, The Ohio State University, Pennsylvania State University, Shanghai Astronomical Observatory, United Kingdom Participation Group, Universidad Nacional Aut\'onoma de M\'exico, University of Arizona, University of Colorado Boulder, University of Oxford, University of Portsmouth, University of Utah, University of Virginia, University of Washington, University of Wisconsin, Vanderbilt University, and Yale University.




\bibliographystyle{mnras}
\bibliography{pa_ref} 

\begin{thebibliography}{}
\makeatletter
\relax
\def\mn@urlcharsother{\let\do\@makeother \do\$\do\&\do\#\do\^\do\_\do\%\do\~}
\def\mn@doi{\begingroup\mn@urlcharsother \@ifnextchar [ {\mn@doi@}
  {\mn@doi@[]}}
\def\mn@doi@[#1]#2{\def\@tempa{#1}\ifx\@tempa\@empty \href
  {http://dx.doi.org/#2} {doi:#2}\else \href {http://dx.doi.org/#2} {#1}\fi
  \endgroup}
\def\mn@eprint#1#2{\mn@eprint@#1:#2::\@nil}
\def\mn@eprint@arXiv#1{\href {http://arxiv.org/abs/#1} {{\tt arXiv:#1}}}
\def\mn@eprint@dblp#1{\href {http://dblp.uni-trier.de/rec/bibtex/#1.xml}
  {dblp:#1}}
\def\mn@eprint@#1:#2:#3:#4\@nil{\def\@tempa {#1}\def\@tempb {#2}\def\@tempc
  {#3}\ifx \@tempc \@empty \let \@tempc \@tempb \let \@tempb \@tempa \fi \ifx
  \@tempb \@empty \def\@tempb {arXiv}\fi \@ifundefined
  {mn@eprint@\@tempb}{\@tempb:\@tempc}{\expandafter \expandafter \csname
  mn@eprint@\@tempb\endcsname \expandafter{\@tempc}}}

\bibitem[\protect\citeauthoryear{{Artale}, {Zehavi}, {Contreras}  \&
  {Norberg}}{{Artale} et~al.}{2018}]{artale2018}
{Artale} M.~C.,  {Zehavi} I.,  {Contreras} S.,   {Norberg} P.,  2018, preprint,
  \href {http://adsabs.harvard.edu/abs/2018arXiv180506938A} {} (\mn@eprint
  {arXiv} {1805.06938})

\bibitem[\protect\citeauthoryear{{Berlind} et~al.,}{{Berlind}
  et~al.}{2003}]{berlind2003}
{Berlind} A.~A.,  et~al., 2003, \mn@doi [\apj] {10.1086/376517}, \href
  {http://adsabs.harvard.edu/abs/2003ApJ...593....1B} {593, 1}

\bibitem[\protect\citeauthoryear{{Blanton} et~al.,}{{Blanton}
  et~al.}{2005}]{blanton2005}
{Blanton} M.~R.,  et~al., 2005, \mn@doi [\aj] {10.1086/429803}, \href
  {http://adsabs.harvard.edu/abs/2005AJ....129.2562B} {129, 2562}

\bibitem[\protect\citeauthoryear{{Blanton} et~al.,}{{Blanton}
  et~al.}{2017}]{blanton2017}
{Blanton} M.~R.,  et~al., 2017, \mn@doi [\aj] {10.3847/1538-3881/aa7567}, \href
  {http://adsabs.harvard.edu/abs/2017AJ....154...28B} {154, 28}

\bibitem[\protect\citeauthoryear{{Bond}, {Cole}, {Efstathiou}  \&
  {Kaiser}}{{Bond} et~al.}{1991}]{bond1991}
{Bond} J.~R.,  {Cole} S.,  {Efstathiou} G.,   {Kaiser} N.,  1991, \mn@doi
  [\apj] {10.1086/170520}, \href
  {http://adsabs.harvard.edu/abs/1991ApJ...379..440B} {379, 440}

\bibitem[\protect\citeauthoryear{{Borzyszkowski}, {Porciani},
  {Romano-D{\'{\i}}az}  \& {Garaldi}}{{Borzyszkowski} et~al.}{2017}]{ZOMGI}
{Borzyszkowski} M.,  {Porciani} C.,  {Romano-D{\'{\i}}az} E.,   {Garaldi} E.,
  2017, \mn@doi [\mnras] {10.1093/mnras/stx873}, \href
  {http://adsabs.harvard.edu/abs/2017MNRAS.469..594B} {469, 594}

\bibitem[\protect\citeauthoryear{{Brouwer} et~al.,}{{Brouwer}
  et~al.}{2016}]{brouwer2016}
{Brouwer} M.~M.,  et~al., 2016, \mn@doi [\mnras] {10.1093/mnras/stw1602}, \href
  {http://adsabs.harvard.edu/abs/2016MNRAS.462.4451B} {462, 4451}

\bibitem[\protect\citeauthoryear{{Bryant} et~al.,}{{Bryant}
  et~al.}{2015}]{bryant2015}
{Bryant} J.~J.,  et~al., 2015, \mn@doi [\mnras] {10.1093/mnras/stu2635}, \href
  {http://adsabs.harvard.edu/abs/2015MNRAS.447.2857B} {447, 2857}

\bibitem[\protect\citeauthoryear{{Bundy} et~al.,}{{Bundy}
  et~al.}{2015}]{bundy2015}
{Bundy} K.,  et~al., 2015, \mn@doi [\apj] {10.1088/0004-637X/798/1/7}, \href
  {http://adsabs.harvard.edu/abs/2015ApJ...798....7B} {798, 7}

\bibitem[\protect\citeauthoryear{Campbell, van~den Bosch, Hearin, Padmanabhan,
  Berlind, Mo, Tinker  \& Yang}{Campbell et~al.}{2015}]{campbell2015}
Campbell D.,  van~den Bosch F.~C.,  Hearin A.,  Padmanabhan N.,  Berlind A.,
  Mo H.~J.,  Tinker J.,   Yang X.,  2015, \mn@doi [\mnras]
  {10.1093/mnras/stv1091}, 452, 444

\bibitem[\protect\citeauthoryear{{Cappellari}}{{Cappellari}}{2017}]{cappellari2017}
{Cappellari} M.,  2017, \mn@doi [\mnras] {10.1093/mnras/stw3020}, \href
  {http://adsabs.harvard.edu/abs/2017MNRAS.466..798C} {466, 798}

\bibitem[\protect\citeauthoryear{{Cappellari} \& {Copin}}{{Cappellari} \&
  {Copin}}{2003}]{cappellari2003}
{Cappellari} M.,  {Copin} Y.,  2003, \mn@doi [\mnras]
  {10.1046/j.1365-8711.2003.06541.x}, \href
  {http://adsabs.harvard.edu/abs/2003MNRAS.342..345C} {342, 345}

\bibitem[\protect\citeauthoryear{{Cappellari} \& {Emsellem}}{{Cappellari} \&
  {Emsellem}}{2004}]{cappellari2004}
{Cappellari} M.,  {Emsellem} E.,  2004, \mn@doi [\pasp] {10.1086/381875}, \href
  {http://adsabs.harvard.edu/abs/2004PASP..116..138C} {116, 138}

\bibitem[\protect\citeauthoryear{{Cappellari} et~al.,}{{Cappellari}
  et~al.}{2011a}]{atlas3d}
{Cappellari} M.,  et~al., 2011a, \mn@doi [\mnras]
  {10.1111/j.1365-2966.2010.18174.x}, \href
  {http://adsabs.harvard.edu/abs/2011MNRAS.413..813C} {413, 813}

\bibitem[\protect\citeauthoryear{{Cappellari} et~al.,}{{Cappellari}
  et~al.}{2011b}]{cappellari2011}
{Cappellari} M.,  et~al., 2011b, \mn@doi [\mnras]
  {10.1111/j.1365-2966.2011.18600.x}, \href
  {http://adsabs.harvard.edu/abs/2011MNRAS.416.1680C} {416, 1680}

\bibitem[\protect\citeauthoryear{{Cautun} \& {van de Weygaert}}{{Cautun} \&
  {van de Weygaert}}{2011}]{cautun2011}
{Cautun} M.~C.,  {van de Weygaert} R.,  2011, {The DTFE public software: The
  Delaunay Tessellation Field Estimator code}, Astrophysics Source Code Library
  (\mn@eprint {ascl} {1105.003})

\bibitem[\protect\citeauthoryear{{Chen} et~al.,}{{Chen}
  et~al.}{2016}]{chen2016}
{Chen} Y.-M.,  et~al., 2016, \mn@doi [Nature Communications]
  {10.1038/ncomms13269}, \href
  {https://ui.adsabs.harvard.edu/#abs/2016NatCo...713269C} {7}

\bibitem[\protect\citeauthoryear{{Conroy}, {Wechsler}  \& {Kravtsov}}{{Conroy}
  et~al.}{2006}]{conroy2006}
{Conroy} C.,  {Wechsler} R.~H.,   {Kravtsov} A.~V.,  2006, \mn@doi [\apj]
  {10.1086/503602}, \href {http://adsabs.harvard.edu/abs/2006ApJ...647..201C}
  {647, 201}

\bibitem[\protect\citeauthoryear{{Correa}, {Schaye}, {van de Voort}, {Duffy}
  \& {Wyithe}}{{Correa} et~al.}{2018}]{correa2018}
{Correa} C.~A.,  {Schaye} J.,  {van de Voort} F.,  {Duffy} A.~R.,   {Wyithe} J.
  S.~B.,  2018, \mn@doi [\mnras] {10.1093/mnras/sty871}, \href
  {https://ui.adsabs.harvard.edu/#abs/2018MNRAS.tmp..846C} {p.~846}

\bibitem[\protect\citeauthoryear{{Cortese} et~al.,}{{Cortese}
  et~al.}{2016}]{cortese2016}
{Cortese} L.,  et~al., 2016, \mn@doi [\mnras] {10.1093/mnras/stw1891}, \href
  {http://adsabs.harvard.edu/abs/2016MNRAS.463..170C} {463, 170}

\bibitem[\protect\citeauthoryear{{Croom} et~al.,}{{Croom}
  et~al.}{2012}]{croom2012}
{Croom} S.~M.,  et~al., 2012, \mn@doi [\mnras]
  {10.1111/j.1365-2966.2011.20365.x}, \href
  {http://adsabs.harvard.edu/abs/2012MNRAS.421..872C} {421, 872}

\bibitem[\protect\citeauthoryear{{Croton}, {Gao}  \& {White}}{{Croton}
  et~al.}{2007}]{croton2007}
{Croton} D.~J.,  {Gao} L.,   {White} S. D.~M.,  2007, \mn@doi [\mnras]
  {10.1111/j.1365-2966.2006.11230.x}, \href
  {https://ui.adsabs.harvard.edu/#abs/2007MNRAS.374.1303C} {374, 1303}

\bibitem[\protect\citeauthoryear{{Dalal}, {White}, {Bond}  \&
  {Shirokov}}{{Dalal} et~al.}{2008}]{dalal2008}
{Dalal} N.,  {White} M.,  {Bond} J.~R.,   {Shirokov} A.,  2008, \mn@doi [\apj]
  {10.1086/591512}, \href
  {https://ui.adsabs.harvard.edu/#abs/2008ApJ...687...12D} {687, 12}

\bibitem[\protect\citeauthoryear{Davis \& Bureau}{Davis \&
  Bureau}{2016}]{davis2016}
Davis T.~A.,  Bureau M.,  2016, \mn@doi [\mnras] {10.1093/mnras/stv2998}, 457,
  272

\bibitem[\protect\citeauthoryear{{Davis} et~al.,}{{Davis}
  et~al.}{2011}]{davis2011a}
{Davis} T.~A.,  et~al., 2011, \mn@doi [\mnras]
  {10.1111/j.1365-2966.2011.19355.x}, \href
  {http://adsabs.harvard.edu/abs/2011MNRAS.417..882D} {417, 882}

\bibitem[\protect\citeauthoryear{{Driver} et~al.,}{{Driver}
  et~al.}{2009}]{driver2009}
{Driver} S.~P.,  et~al., 2009, \mn@doi [Astronomy and Geophysics]
  {10.1111/j.1468-4004.2009.50512.x}, \href
  {http://adsabs.harvard.edu/abs/2009A%26G....50e..12D} {50, 5.12}

\bibitem[\protect\citeauthoryear{{Driver} et~al.,}{{Driver}
  et~al.}{2011}]{driver2011}
{Driver} S.~P.,  et~al., 2011, \mn@doi [\mnras]
  {10.1111/j.1365-2966.2010.18188.x}, \href
  {http://adsabs.harvard.edu/abs/2011MNRAS.413..971D} {413, 971}

\bibitem[\protect\citeauthoryear{{Drory} et~al.,}{{Drory}
  et~al.}{2015}]{drory2015}
{Drory} N.,  et~al., 2015, \mn@doi [\aj] {10.1088/0004-6256/149/2/77}, \href
  {http://adsabs.harvard.edu/abs/2015AJ....149...77D} {149, 77}

\bibitem[\protect\citeauthoryear{{Eardley} et~al.,}{{Eardley}
  et~al.}{2015}]{eardley2015}
{Eardley} E.,  et~al., 2015, \mn@doi [\mnras] {10.1093/mnras/stv237}, \href
  {https://ui.adsabs.harvard.edu/#abs/2015MNRAS.448.3665E} {448, 3665}

\bibitem[\protect\citeauthoryear{{Falc{\'o}n-Barroso},
  {S{\'a}nchez-Bl{\'a}zquez}, {Vazdekis}, {Ricciardelli}, {Cardiel}, {Cenarro},
  {Gorgas}  \& {Peletier}}{{Falc{\'o}n-Barroso} et~al.}{2011}]{falcon2011}
{Falc{\'o}n-Barroso} J.,  {S{\'a}nchez-Bl{\'a}zquez} P.,  {Vazdekis} A.,
  {Ricciardelli} E.,  {Cardiel} N.,  {Cenarro} A.~J.,  {Gorgas} J.,
  {Peletier} R.~F.,  2011, \mn@doi [\aap] {10.1051/0004-6361/201116842}, \href
  {http://adsabs.harvard.edu/abs/2011A%26A...532A..95F} {532, A95}

\bibitem[\protect\citeauthoryear{{Gao} \& {White}}{{Gao} \&
  {White}}{2007}]{gao2007}
{Gao} L.,  {White} S.~D.~M.,  2007, \mn@doi [\mnras]
  {10.1111/j.1745-3933.2007.00292.x}, \href
  {http://adsabs.harvard.edu/abs/2007MNRAS.377L...5G} {377, L5}

\bibitem[\protect\citeauthoryear{{Gao}, {Springel}  \& {White}}{{Gao}
  et~al.}{2005}]{gao2005}
{Gao} L.,  {Springel} V.,   {White} S.~D.~M.,  2005, \mn@doi [\mnras]
  {10.1111/j.1745-3933.2005.00084.x}, \href
  {http://adsabs.harvard.edu/abs/2005MNRAS.363L..66G} {363, L66}

\bibitem[\protect\citeauthoryear{{Greene} et~al.,}{{Greene}
  et~al.}{2018}]{greene2018}
{Greene} J.~E.,  et~al., 2018, \mn@doi [\apj] {10.3847/1538-4357/aa9bde}, \href
  {http://ukads.nottingham.ac.uk/abs/2018ApJ...852...36G} {852, 36}

\bibitem[\protect\citeauthoryear{{Gunn} et~al.,}{{Gunn}
  et~al.}{2006}]{gunn2006}
{Gunn} J.~E.,  et~al., 2006, \mn@doi [\aj] {10.1086/500975}, \href
  {http://adsabs.harvard.edu/abs/2006AJ....131.2332G} {131, 2332}

\bibitem[\protect\citeauthoryear{{Hahn}, {Porciani}, {Dekel}  \&
  {Carollo}}{{Hahn} et~al.}{2009}]{hahn2009}
{Hahn} O.,  {Porciani} C.,  {Dekel} A.,   {Carollo} C.~M.,  2009, \mn@doi
  [\mnras] {10.1111/j.1365-2966.2009.15271.x}, \href
  {http://adsabs.harvard.edu/abs/2009MNRAS.398.1742H} {398, 1742}

\bibitem[\protect\citeauthoryear{{Hernquist}}{{Hernquist}}{1990}]{hernquist1990}
{Hernquist} L.,  1990, \mn@doi [\apj] {10.1086/168845}, \href
  {http://adsabs.harvard.edu/abs/1990ApJ...356..359H} {356, 359}

\bibitem[\protect\citeauthoryear{{Jackson}}{{Jackson}}{1972}]{jackson1972}
{Jackson} J.~C.,  1972, \mn@doi [\mnras] {10.1093/mnras/156.1.1P}, \href
  {https://ui.adsabs.harvard.edu/#abs/1972MNRAS.156P...1J} {156, 1P}

\bibitem[\protect\citeauthoryear{{Jin} et~al.,}{{Jin} et~al.}{2016}]{jin2016}
{Jin} Y.,  et~al., 2016, \mn@doi [\mnras] {10.1093/mnras/stw2055}, \href
  {http://adsabs.harvard.edu/abs/2016MNRAS.463..913J} {463, 913}

\bibitem[\protect\citeauthoryear{{Jing}, {Mo}  \& {B{\"o}rner}}{{Jing}
  et~al.}{1998}]{jing1998}
{Jing} Y.~P.,  {Mo} H.~J.,   {B{\"o}rner} G.,  1998, \mn@doi [\apj]
  {10.1086/305209}, \href {http://adsabs.harvard.edu/abs/1998ApJ...494....1J}
  {494, 1}

\bibitem[\protect\citeauthoryear{{Jing}, {Suto}  \& {Mo}}{{Jing}
  et~al.}{2007}]{jing2007}
{Jing} Y.~P.,  {Suto} Y.,   {Mo} H.~J.,  2007, \mn@doi [\apj] {10.1086/511130},
  \href {http://adsabs.harvard.edu/abs/2007ApJ...657..664J} {657, 664}

\bibitem[\protect\citeauthoryear{{Kaiser}}{{Kaiser}}{1987}]{kaiser1987}
{Kaiser} N.,  1987, \mn@doi [\mnras] {10.1093/mnras/227.1.1}, \href
  {https://ui.adsabs.harvard.edu/#abs/1987MNRAS.227....1K} {227, 1}

\bibitem[\protect\citeauthoryear{{Krajnovi{\'c}}, {Cappellari}, {de Zeeuw}  \&
  {Copin}}{{Krajnovi{\'c}} et~al.}{2006}]{krajnovic2006}
{Krajnovi{\'c}} D.,  {Cappellari} M.,  {de Zeeuw} P.~T.,   {Copin} Y.,  2006,
  \mn@doi [\mnras] {10.1111/j.1365-2966.2005.09902.x}, \href
  {http://adsabs.harvard.edu/abs/2006MNRAS.366..787K} {366, 787}

\bibitem[\protect\citeauthoryear{{Kraljic} et~al.,}{{Kraljic}
  et~al.}{2018}]{kraljic2018}
{Kraljic} K.,  et~al., 2018, \mn@doi [\mnras] {10.1093/mnras/stx2638}, \href
  {http://adsabs.harvard.edu/abs/2018MNRAS.474..547K} {474, 547}

\bibitem[\protect\citeauthoryear{{Kravtsov}, {Berlind}, {Wechsler}, {Klypin},
  {Gottl{\"o}ber}, {Allgood}  \& {Primack}}{{Kravtsov}
  et~al.}{2004}]{kravtsov2004}
{Kravtsov} A.~V.,  {Berlind} A.~A.,  {Wechsler} R.~H.,  {Klypin} A.~A.,
  {Gottl{\"o}ber} S.,  {Allgood} B.,   {Primack} J.~R.,  2004, \mn@doi [\apj]
  {10.1086/420959}, \href {http://adsabs.harvard.edu/abs/2004ApJ...609...35K}
  {609, 35}

\bibitem[\protect\citeauthoryear{Lacerna \& Padilla}{Lacerna \&
  Padilla}{2011}]{lacerna2011}
Lacerna I.,  Padilla N.,  2011, \mn@doi [\mnras]
  {10.1111/j.1365-2966.2010.17988.x}, 412, 1283

\bibitem[\protect\citeauthoryear{{Lacerna} \& {Padilla}}{{Lacerna} \&
  {Padilla}}{2012}]{lacerna2012}
{Lacerna} I.,  {Padilla} N.,  2012, \mn@doi [\mnras]
  {10.1111/j.1745-3933.2012.01316.x}, \href
  {http://adsabs.harvard.edu/abs/2012MNRAS.426L..26L} {426, L26}

\bibitem[\protect\citeauthoryear{{Lagos}, {Padilla}, {Davis}, {Lacey}, {Baugh},
  {Gonzalez-Perez}, {Zwaan}  \& {Contreras}}{{Lagos} et~al.}{2015}]{lagos2015}
{Lagos} C.~d.~P.,  {Padilla} N.~D.,  {Davis} T.~A.,  {Lacey} C.~G.,  {Baugh}
  C.~M.,  {Gonzalez-Perez} V.,  {Zwaan} M.~A.,   {Contreras} S.,  2015, \mn@doi
  [\mnras] {10.1093/mnras/stu2763}, \href
  {http://adsabs.harvard.edu/abs/2015MNRAS.448.1271L} {448, 1271}

\bibitem[\protect\citeauthoryear{{Lagos}, {Theuns}, {Stevens}, {Cortese},
  {Padilla}, {Davis}, {Contreras}  \& {Croton}}{{Lagos}
  et~al.}{2017}]{lagos2017}
{Lagos} C.~d.~P.,  {Theuns} T.,  {Stevens} A.~R.~H.,  {Cortese} L.,  {Padilla}
  N.~D.,  {Davis} T.~A.,  {Contreras} S.,   {Croton} D.,  2017, \mn@doi
  [\mnras] {10.1093/mnras/stw2610}, \href
  {http://ukads.nottingham.ac.uk/abs/2017MNRAS.464.3850L} {464, 3850}

\bibitem[\protect\citeauthoryear{{Lagos}, {Schaye}, {Bah{\'e}}, {Van de Sande},
  {Kay}, {Barnes}, {Davis}  \& {Dalla Vecchia}}{{Lagos}
  et~al.}{2018}]{lagos2018}
{Lagos} C.~d.~P.,  {Schaye} J.,  {Bah{\'e}} Y.,  {Van de Sande} J.,  {Kay}
  S.~T.,  {Barnes} D.,  {Davis} T.~A.,   {Dalla Vecchia} C.,  2018, \mn@doi
  [\mnras] {10.1093/mnras/sty489}, \href
  {http://adsabs.harvard.edu/abs/2018MNRAS.tmp..476L} {}

\bibitem[\protect\citeauthoryear{{Law} et~al.,}{{Law}
  et~al.}{2015}]{law2015obs}
{Law} D.~R.,  et~al., 2015, \mn@doi [\aj] {10.1088/0004-6256/150/1/19}, \href
  {http://adsabs.harvard.edu/abs/2015AJ....150...19L} {150, 19}

\bibitem[\protect\citeauthoryear{{Law} et~al.,}{{Law}
  et~al.}{2016}]{law2016drp}
{Law} D.~R.,  et~al., 2016, \mn@doi [\aj] {10.3847/0004-6256/152/4/83}, \href
  {http://adsabs.harvard.edu/abs/2016AJ....152...83L} {152, 83}

\bibitem[\protect\citeauthoryear{{Lehmann}, {Mao}, {Becker}, {Skillman}  \&
  {Wechsler}}{{Lehmann} et~al.}{2017}]{lehmann2017}
{Lehmann} B.~V.,  {Mao} Y.-Y.,  {Becker} M.~R.,  {Skillman} S.~W.,   {Wechsler}
  R.~H.,  2017, \mn@doi [\apj] {10.3847/1538-4357/834/1/37}, \href
  {http://adsabs.harvard.edu/abs/2017ApJ...834...37L} {834, 37}

\bibitem[\protect\citeauthoryear{{Lim}, {Mo}, {Wang}  \& {Yang}}{{Lim}
  et~al.}{2016}]{lim2015}
{Lim} S.~H.,  {Mo} H.~J.,  {Wang} H.,   {Yang} X.,  2016, \mn@doi [\mnras]
  {10.1093/mnras/stv2282}, \href
  {http://adsabs.harvard.edu/abs/2016MNRAS.455..499L} {455, 499}

\bibitem[\protect\citeauthoryear{{Lintott} et~al.,}{{Lintott}
  et~al.}{2008}]{lintott2008}
{Lintott} C.~J.,  et~al., 2008, \mn@doi [\mnras]
  {10.1111/j.1365-2966.2008.13689.x}, \href
  {http://ukads.nottingham.ac.uk/abs/2008MNRAS.389.1179L} {389, 1179}

\bibitem[\protect\citeauthoryear{{Liu}, {Gerke}, {Wechsler}, {Behroozi}  \&
  {Busha}}{{Liu} et~al.}{2011}]{liu2011}
{Liu} L.,  {Gerke} B.~F.,  {Wechsler} R.~H.,  {Behroozi} P.~S.,   {Busha}
  M.~T.,  2011, \mn@doi [\apj] {10.1088/0004-637X/733/1/62}, \href
  {http://adsabs.harvard.edu/abs/2011ApJ...733...62L} {733, 62}

\bibitem[\protect\citeauthoryear{{Malavasi} et~al.,}{{Malavasi}
  et~al.}{2017}]{malavasi2017}
{Malavasi} N.,  et~al., 2017, \mn@doi [\mnras] {10.1093/mnras/stw2864}, \href
  {https://ui.adsabs.harvard.edu/#abs/2017MNRAS.465.3817M} {465, 3817}

\bibitem[\protect\citeauthoryear{{Matthee}, {Schaye}, {Crain}, {Schaller},
  {Bower}  \& {Theuns}}{{Matthee} et~al.}{2017}]{matthee2017}
{Matthee} J.,  {Schaye} J.,  {Crain} R.~A.,  {Schaller} M.,  {Bower} R.,
  {Theuns} T.,  2017, \mn@doi [\mnras] {10.1093/mnras/stw2884}, \href
  {http://adsabs.harvard.edu/abs/2017MNRAS.465.2381M} {465, 2381}

\bibitem[\protect\citeauthoryear{{Mo} \& {White}}{{Mo} \&
  {White}}{1996}]{mo1996}
{Mo} H.~J.,  {White} S.~D.~M.,  1996, \mn@doi [\mnras]
  {10.1093/mnras/282.2.347}, \href
  {http://adsabs.harvard.edu/abs/1996MNRAS.282..347M} {282, 347}

\bibitem[\protect\citeauthoryear{{Musso}, {Cadiou}, {Pichon}, {Codis},
  {Kraljic}  \& {Dubois}}{{Musso} et~al.}{2018}]{musso2018}
{Musso} M.,  {Cadiou} C.,  {Pichon} C.,  {Codis} S.,  {Kraljic} K.,   {Dubois}
  Y.,  2018, \mn@doi [\mnras] {10.1093/mnras/sty191}, \href
  {http://adsabs.harvard.edu/abs/2018MNRAS.476.4877M} {476, 4877}

\bibitem[\protect\citeauthoryear{Nelson, Vogelsberger, Genel, Sijacki, Kereš,
  Springel  \& Hernquist}{Nelson et~al.}{2013}]{nelson2013}
Nelson D.,  Vogelsberger M.,  Genel S.,  Sijacki D.,  Kereš D.,  Springel V.,
   Hernquist L.,  2013, \mn@doi [\mnras] {10.1093/mnras/sts595}, 429, 3353

\bibitem[\protect\citeauthoryear{{Padilla}, {Salazar-Albornoz}, {Contreras},
  {Cora}  \& {Ruiz}}{{Padilla} et~al.}{2014}]{padilla2014}
{Padilla} N.~D.,  {Salazar-Albornoz} S.,  {Contreras} S.,  {Cora} S.~A.,
  {Ruiz} A.~N.,  2014, \mn@doi [\mnras] {10.1093/mnras/stu1321}, \href
  {https://ui.adsabs.harvard.edu/#abs/2014MNRAS.443.2801P} {443, 2801}

\bibitem[\protect\citeauthoryear{{Peacock} \& {Smith}}{{Peacock} \&
  {Smith}}{2000}]{peacock2000}
{Peacock} J.~A.,  {Smith} R.~E.,  2000, \mn@doi [\mnras]
  {10.1046/j.1365-8711.2000.03779.x}, \href
  {http://adsabs.harvard.edu/abs/2000MNRAS.318.1144P} {318, 1144}

\bibitem[\protect\citeauthoryear{{Pichon}, {Pogosyan}, {Kimm}, {Slyz},
  {Devriendt}  \& {Dubois}}{{Pichon} et~al.}{2011}]{pichon2011}
{Pichon} C.,  {Pogosyan} D.,  {Kimm} T.,  {Slyz} A.,  {Devriendt} J.,
  {Dubois} Y.,  2011, \mn@doi [\mnras] {10.1111/j.1365-2966.2011.19640.x},
  \href {https://ui.adsabs.harvard.edu/#abs/2011MNRAS.418.2493P} {418, 2493}

\bibitem[\protect\citeauthoryear{{Planck Collaboration} et~al.,}{{Planck
  Collaboration} et~al.}{2016}]{planck2016}
{Planck Collaboration} et~al., 2016, \mn@doi [\aap]
  {10.1051/0004-6361/201525830}, \href
  {http://adsabs.harvard.edu/abs/2016A%26A...594A..13P} {594, A13}

\bibitem[\protect\citeauthoryear{{Press} \& {Schechter}}{{Press} \&
  {Schechter}}{1974}]{press1974}
{Press} W.~H.,  {Schechter} P.,  1974, \mn@doi [\apj] {10.1086/152650}, \href
  {http://adsabs.harvard.edu/abs/1974ApJ...187..425P} {187, 425}

\bibitem[\protect\citeauthoryear{{Reddick}, {Wechsler}, {Tinker}  \&
  {Behroozi}}{{Reddick} et~al.}{2013}]{reddick2013}
{Reddick} R.~M.,  {Wechsler} R.~H.,  {Tinker} J.~L.,   {Behroozi} P.~S.,  2013,
  \mn@doi [\apj] {10.1088/0004-637X/771/1/30}, \href
  {https://ui.adsabs.harvard.edu/#abs/2013ApJ...771...30R} {771, 30}

\bibitem[\protect\citeauthoryear{{Rodriguez}, {Merch{\'a}n}  \&
  {Sgr{\'o}}}{{Rodriguez} et~al.}{2015}]{rodriguez2015}
{Rodriguez} F.,  {Merch{\'a}n} M.,   {Sgr{\'o}} M.~A.,  2015, \mn@doi [\aap]
  {10.1051/0004-6361/201525798}, \href
  {http://adsabs.harvard.edu/abs/2015A%26A...580A..86R} {580, A86}

\bibitem[\protect\citeauthoryear{{S{\'a}nchez-Bl{\'a}zquez}
  et~al.,}{{S{\'a}nchez-Bl{\'a}zquez} et~al.}{2006}]{sanchez2006}
{S{\'a}nchez-Bl{\'a}zquez} P.,  et~al., 2006, \mn@doi [\mnras]
  {10.1111/j.1365-2966.2006.10699.x}, \href
  {http://adsabs.harvard.edu/abs/2006MNRAS.371..703S} {371, 703}

\bibitem[\protect\citeauthoryear{{Sarzi} et~al.,}{{Sarzi}
  et~al.}{2006}]{sarzi2006}
{Sarzi} M.,  et~al., 2006, \mn@doi [\mnras] {10.1111/j.1365-2966.2005.09839.x},
  \href {http://adsabs.harvard.edu/abs/2006MNRAS.366.1151S} {366, 1151}

\bibitem[\protect\citeauthoryear{{Schaap} \& {van de Weygaert}}{{Schaap} \&
  {van de Weygaert}}{2000}]{schaap2000}
{Schaap} W.~E.,  {van de Weygaert} R.,  2000, \aap, \href
  {https://ui.adsabs.harvard.edu/#abs/2000A&A...363L..29S} {363, L29}

\bibitem[\protect\citeauthoryear{{Schaye} et~al.,}{{Schaye}
  et~al.}{2015}]{EAGLE2015}
{Schaye} J.,  et~al., 2015, \mn@doi [\mnras] {10.1093/mnras/stu2058}, \href
  {http://adsabs.harvard.edu/abs/2015MNRAS.446..521S} {446, 521}

\bibitem[\protect\citeauthoryear{{Sheth} \& {Tormen}}{{Sheth} \&
  {Tormen}}{1999}]{sheth1999}
{Sheth} R.~K.,  {Tormen} G.,  1999, \mn@doi [\mnras]
  {10.1046/j.1365-8711.1999.02692.x}, \href
  {https://ui.adsabs.harvard.edu/#abs/1999MNRAS.308..119S} {308, 119}

\bibitem[\protect\citeauthoryear{{Shi} et~al.,}{{Shi} et~al.}{2016}]{shi2016}
{Shi} F.,  et~al., 2016, \mn@doi [\apj] {10.3847/1538-4357/833/2/241}, \href
  {https://ui.adsabs.harvard.edu/#abs/2016ApJ...833..241S} {833, 241}

\bibitem[\protect\citeauthoryear{{Smee} et~al.,}{{Smee}
  et~al.}{2013}]{smee2013}
{Smee} S.~A.,  et~al., 2013, \mn@doi [\aj] {10.1088/0004-6256/146/2/32}, \href
  {http://adsabs.harvard.edu/abs/2013AJ....146...32S} {146, 32}

\bibitem[\protect\citeauthoryear{{Sousbie}}{{Sousbie}}{2011}]{sousbie2011a}
{Sousbie} T.,  2011, \mn@doi [\mnras] {10.1111/j.1365-2966.2011.18394.x}, \href
  {http://adsabs.harvard.edu/abs/2011MNRAS.414..350S} {414, 350}

\bibitem[\protect\citeauthoryear{{Sousbie}, {Pichon}  \& {Kawahara}}{{Sousbie}
  et~al.}{2011}]{sousbie2011b}
{Sousbie} T.,  {Pichon} C.,   {Kawahara} H.,  2011, \mn@doi [\mnras]
  {10.1111/j.1365-2966.2011.18395.x}, \href
  {http://adsabs.harvard.edu/abs/2011MNRAS.414..384S} {414, 384}

\bibitem[\protect\citeauthoryear{{Springel} et~al.,}{{Springel}
  et~al.}{2005}]{springel2005}
{Springel} V.,  et~al., 2005, \mn@doi [\nat] {10.1038/nature03597}, \href
  {http://adsabs.harvard.edu/abs/2005Natur.435..629S} {435, 629}

\bibitem[\protect\citeauthoryear{{Stark} et~al.,}{{Stark}
  et~al.}{2018}]{stark2018}
{Stark} D.~V.,  et~al., 2018, in American Astronomical Society, AAS Meeting
  231, id. 210.01.

\bibitem[\protect\citeauthoryear{{Tempel} et~al.,}{{Tempel}
  et~al.}{2014}]{tempel2014}
{Tempel} E.,  et~al., 2014, \mn@doi [\aap] {10.1051/0004-6361/201423585}, \href
  {http://adsabs.harvard.edu/abs/2014A%26A...566A...1T} {566, A1}

\bibitem[\protect\citeauthoryear{{Tojeiro} et~al.,}{{Tojeiro}
  et~al.}{2017}]{tojeiro2017}
{Tojeiro} R.,  et~al., 2017, \mn@doi [\mnras] {10.1093/mnras/stx1466}, \href
  {http://adsabs.harvard.edu/abs/2017MNRAS.470.3720T} {470, 3720}

\bibitem[\protect\citeauthoryear{{Tully} \& {Fisher}}{{Tully} \&
  {Fisher}}{1978}]{tulley1978}
{Tully} R.~B.,  {Fisher} J.~R.,  1978, in {Longair} M.~S.,  {Einasto} J.,  eds,
   Vol. 79, Large Scale Structures in the Universe. p.~31

\bibitem[\protect\citeauthoryear{{Wake} et~al.,}{{Wake}
  et~al.}{2017}]{wake2017}
{Wake} D.~A.,  et~al., 2017, \mn@doi [\aj] {10.3847/1538-3881/aa7ecc}, \href
  {http://adsabs.harvard.edu/abs/2017AJ....154...86W} {154, 86}

\bibitem[\protect\citeauthoryear{{Wang}, {Mo}  \& {Jing}}{{Wang}
  et~al.}{2007}]{wang2007}
{Wang} H.~Y.,  {Mo} H.~J.,   {Jing} Y.~P.,  2007, \mn@doi [\mnras]
  {10.1111/j.1365-2966.2006.11316.x}, \href
  {http://adsabs.harvard.edu/abs/2007MNRAS.375..633W} {375, 633}

\bibitem[\protect\citeauthoryear{{Wang}, {Mo}, {Jing}, {Yang}  \&
  {Wang}}{{Wang} et~al.}{2011}]{wang2011}
{Wang} H.,  {Mo} H.~J.,  {Jing} Y.~P.,  {Yang} X.,   {Wang} Y.,  2011, \mn@doi
  [\mnras] {10.1111/j.1365-2966.2011.18301.x}, \href
  {http://adsabs.harvard.edu/abs/2011MNRAS.413.1973W} {413, 1973}

\bibitem[\protect\citeauthoryear{{Wechsler}, {Zentner}, {Bullock}, {Kravtsov}
  \& {Allgood}}{{Wechsler} et~al.}{2006}]{wechsler2006}
{Wechsler} R.~H.,  {Zentner} A.~R.,  {Bullock} J.~S.,  {Kravtsov} A.~V.,
  {Allgood} B.,  2006, \mn@doi [\apj] {10.1086/507120}, \href
  {https://ui.adsabs.harvard.edu/#abs/2006ApJ...652...71W} {652, 71}

\bibitem[\protect\citeauthoryear{{White} \& {Rees}}{{White} \&
  {Rees}}{1978}]{white1978}
{White} S.~D.~M.,  {Rees} M.~J.,  1978, \mn@doi [\mnras]
  {10.1093/mnras/183.3.341}, \href
  {http://adsabs.harvard.edu/abs/1978MNRAS.183..341W} {183, 341}

\bibitem[\protect\citeauthoryear{{Yan} et~al.,}{{Yan}
  et~al.}{2016a}]{yan2016spec}
{Yan} R.,  et~al., 2016a, \mn@doi [\aj] {10.3847/0004-6256/151/1/8}, \href
  {http://adsabs.harvard.edu/abs/2016AJ....151....8Y} {151, 8}

\bibitem[\protect\citeauthoryear{{Yan} et~al.,}{{Yan}
  et~al.}{2016b}]{yan2016obs}
{Yan} R.,  et~al., 2016b, \mn@doi [\aj] {10.3847/0004-6256/152/6/197}, \href
  {http://adsabs.harvard.edu/abs/2016AJ....152..197Y} {152, 197}

\bibitem[\protect\citeauthoryear{{Yang}, {Mo}, {van den Bosch}, {Pasquali},
  {Li}  \& {Barden}}{{Yang} et~al.}{2007}]{yang2007}
{Yang} X.,  {Mo} H.~J.,  {van den Bosch} F.~C.,  {Pasquali} A.,  {Li} C.,
  {Barden} M.,  2007, \mn@doi [\apj] {10.1086/522027}, \href
  {http://adsabs.harvard.edu/abs/2007ApJ...671..153Y} {671, 153}

\bibitem[\protect\citeauthoryear{{Yang}, {Mo}  \& {van den Bosch}}{{Yang}
  et~al.}{2008}]{yang2008}
{Yang} X.,  {Mo} H.~J.,   {van den Bosch} F.~C.,  2008, \mn@doi [\apj]
  {10.1086/528954}, \href {http://adsabs.harvard.edu/abs/2008ApJ...676..248Y}
  {676, 248}

\bibitem[\protect\citeauthoryear{{Yang}, {Mo}  \& {van den Bosch}}{{Yang}
  et~al.}{2009}]{yang2009}
{Yang} X.,  {Mo} H.~J.,   {van den Bosch} F.~C.,  2009, \mn@doi [\apj]
  {10.1088/0004-637X/695/2/900}, \href
  {http://ukads.nottingham.ac.uk/abs/2009ApJ...695..900Y} {695, 900}

\bibitem[\protect\citeauthoryear{{Zehavi} et~al.,}{{Zehavi}
  et~al.}{2011}]{zehavi2011}
{Zehavi} I.,  et~al., 2011, \mn@doi [\apj] {10.1088/0004-637X/736/1/59}, \href
  {http://adsabs.harvard.edu/abs/2011ApJ...736...59Z} {736, 59}

\bibitem[\protect\citeauthoryear{{Zehavi}, {Contreras}, {Padilla}, {Smith},
  {Baugh}  \& {Norberg}}{{Zehavi} et~al.}{2018}]{zehavi2018}
{Zehavi} I.,  {Contreras} S.,  {Padilla} N.,  {Smith} N.~J.,  {Baugh} C.~M.,
  {Norberg} P.,  2018, \mn@doi [\apj] {10.3847/1538-4357/aaa54a}, \href
  {http://adsabs.harvard.edu/abs/2018ApJ...853...84Z} {853, 84}

\bibitem[\protect\citeauthoryear{{van de Voort}, {Davis}, {Kere{\v s}},
  {Quataert}, {Fauchaer-Gigu{\`e}re}  \& {Hopkins}}{{van de Voort}
  et~al.}{2015}]{vdvoort2015}
{van de Voort} F.,  {Davis} T.~A.,  {Kere{\v s}} D.,  {Quataert} E.,
  {Fauchaer-Gigu{\`e}re} C.-A.,   {Hopkins} P.~F.,  2015, \mn@doi [\mnras]
  {10.1093/mnras/stv1217}, \href
  {http://adsabs.harvard.edu/abs/2015MNRAS.451.3269V} {451, 3269}

\makeatother
\end{thebibliography}




\appendix
\section{Model velocity maps}
\subsection{Circular velocity}
In this appendix we investigate the typical fitting errors of \texttt{FIT\_KINEMATIC\_PA} for the MaNGA sample. It is an important point to constrain the errors of our PA fits, so we can reliably trust cuts in $\Delta$PA to select galaxies which are significantly kinematically misaligned and hence have had external interaction.

Errors using the \texttt{FIT\_KINEMATIC\_PA} routine have been previously estimated for molecular gas velocity fields in ATLAS\textsuperscript{3D} \citep{davis2011a}. Model velocity fields with a known PA were constructed using an empirical galaxy rotation curve and combined with Gaussian noise matched to the signal-to-noise ratio of the data. A typical scatter of $\approx10^{\circ}$ was found due to varying inclination and angular resolution for the velocity fields.

To find the typical error on $\Delta$PA for galaxies in MaNGA, we create model velocity maps for both the stellar and gas components of each MPL-6 observation. In each instance the basic construction of the model follows Section 4 of \citet{krajnovic2006}. Each velocity field comprises of a two-component Hernquist potential which provides a basic circular velocity given by,
\begin{equation}
V_c = \frac{\sqrt{GMr}}{r+r_0}
\end{equation}
where $G$ is the gravitational constant, $M$ is the total mass and $r_0$ is the core radius of each component respectively \citep{hernquist1990}. We use a two-component model to include the relative strengths of both disc and bulge each with distinct effective radii, $R_e$. We fix $r_0$ to be 5 and 15 (units: $arcsec$) and $\sqrt{GM}$ to be 850 and 1500 (units: $km s^{-1} arcsec^{1/2}) $ for the bulge and disc components respectively. These individual components are light weighted by model sersic flux profiles according to,
\begin{equation}
I(r) = I_0 e^{-\left(\frac{R}{R_e}\right)^{n_s}}
\end{equation}
where $I_{0}$ is the peak flux and $n_s$ is the sersic index which is set to 1 and 4 for the disc and spheroidal components respectively. Since we do not have bulge-disc decompositions, we lack individual effective radii for both the bulge and disc components. Instead, we set the bulge and disc effective radii to be 0.5$R_e$ and 1.5$R_e$, where $R_e$ is the effective radius estimated by an elliptical petrosian fit taken from the NSA targeting catalogue introduced in \S\ref{MaNGA_survey}.

\subsection{Calibration}
For each MaNGA galaxy a basic velocity field model is constructed using this template. The axes of the model velocity field are then scaled according to the inclination, $i$, which is estimated from the $b/a$ ratio taken from the NSA catalogue and is also used to scale the fraction of rotational velocity along the line-of-sight. The velocity field for each component, ($j=bulge,disc$), in polar coordinates $(r,\phi)$ is then described by,
\begin{equation}
V(r,\phi) = \frac{I_{j}(r)}{I_{tot}(r)}V_{c}(r)\cos(\phi+\theta_{j})\sin(i)
\end{equation}
where $\theta_j$ is the input kinematic position angle. We set $\theta_{bulge} = \theta_{disc}$ for simplicity, however, we do note that galaxies with more complex orbital motions may increase the typical error. The position angle for both bulge and disc is simply taken to be the opening angle of the galaxy (direction of major axis taken from NSA catalogue). 

In order to imitate a MaNGA observation, the model velocity field is sampled at the spatial resolution of the corresponding IFU bundle and projected into the original shape of the actual observation for H$\alpha$ and stellar maps respectively. Gaussian noise is drawn for each spaxel from a normal distribution with the standard deviation taken from the errors on the actual observation. In addition, these model velocity fields are then Voronoi binned to match the original observation.

\begin{figure}
	\includegraphics[width=\linewidth]{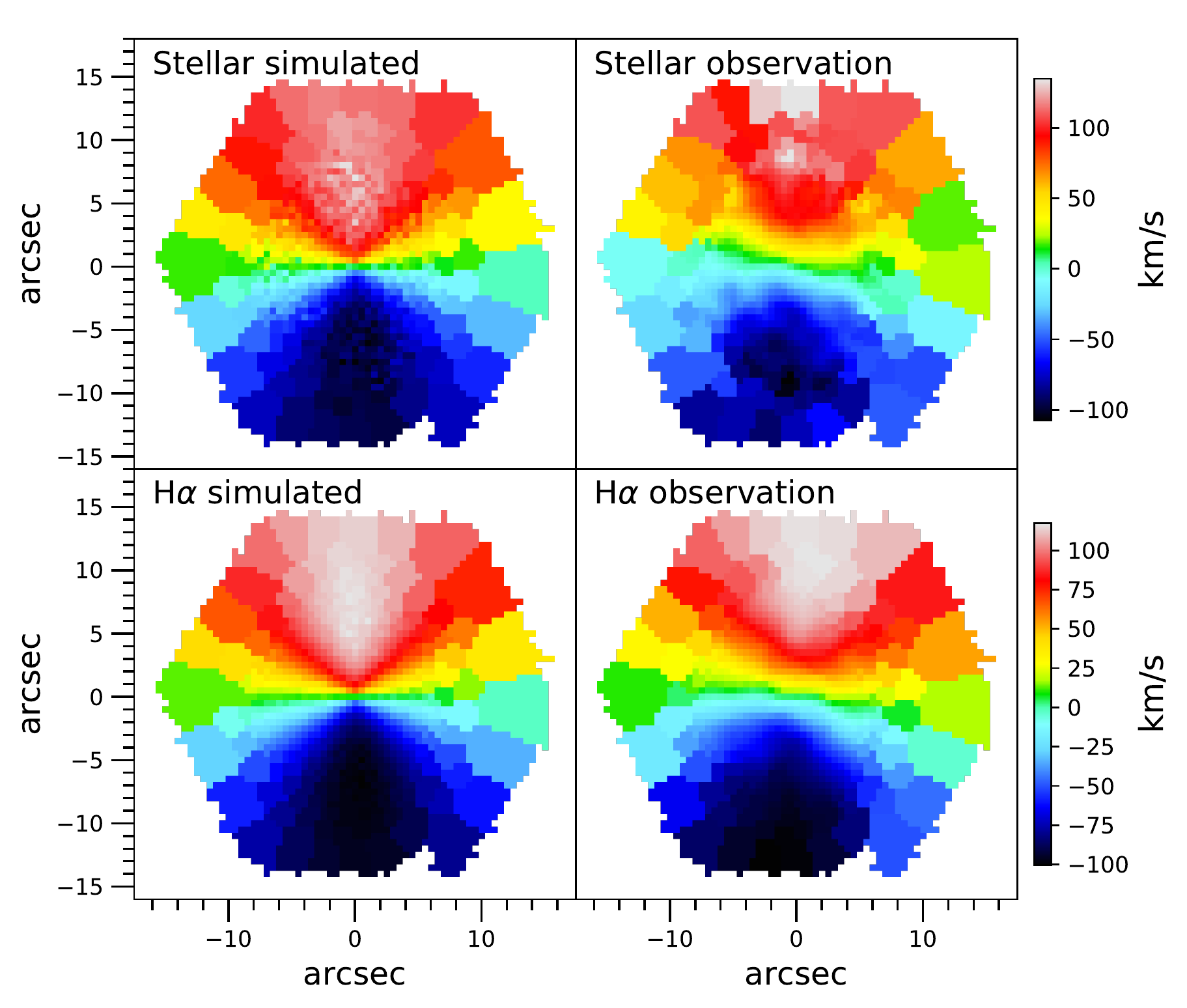}
    \caption{Comparison velocity maps for simulation (left column) and observation (right column). Stellar (H$\alpha$) component velocity maps are shown on the top (bottom) row with the associated velocity colourbars.}
    \label{fig:sim_ifu}
\end{figure}

Example stellar and H$\alpha$ velocity fields generated from these models are shown in Figure \ref{fig:sim_ifu} with comparison to the actual observation. As expected, the model velocity fields frequently recreate observations well but struggle to encompass more complex motion. For this reason, our models should make a reasonable prediction on the typical $\Delta$PA errors intrinsic to MaNGA observations.

\subsection{Typical errors}
\begin{figure}
	\includegraphics[width=\linewidth]{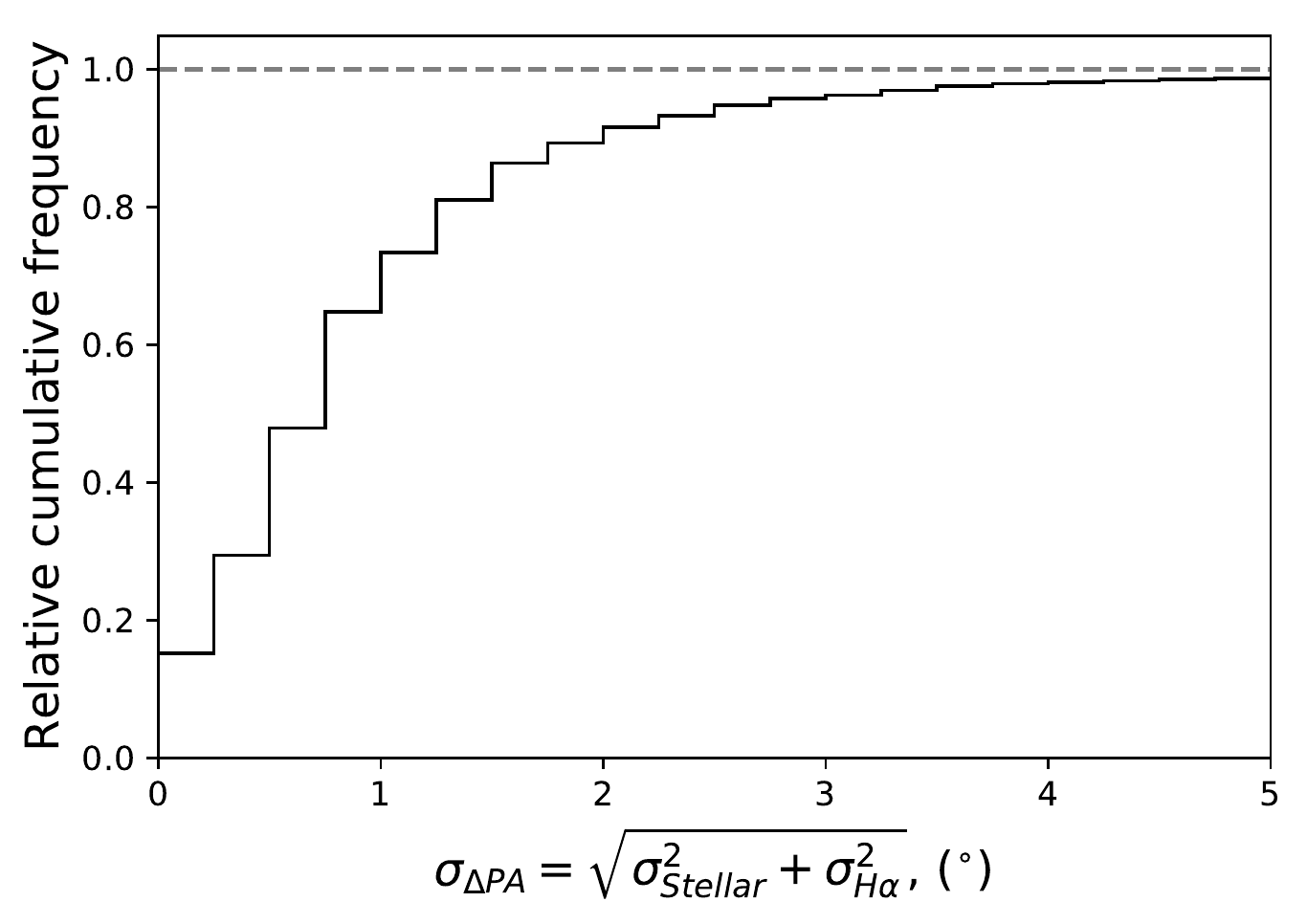}
    \caption{Cumulative histogram of errors for fitting kinematic PA to model maps for the all non-critically flagged MPL-6 galaxy observations.}
    \label{fig:model_errors}
\end{figure}

We construct model velocity fields for all non-critically flagged MPL-6 galaxies, inclusive of the $\Delta$PA sample used in this work. Figure \ref{fig:model_errors} shows the cumulative probability distribution for the range of $0-5^{\circ}$ where the majority of errors fall. We find that \texttt{FIT\_KINEMATIC\_PA} gives a typical combined (stellar and gas) mean error of $1.3^{\circ}$. While this is an underestimation of true $\Delta$PA errors for a sample of galaxies including those with more complex velocity fields, it is indicative that selecting a cut at $\Delta$PA = 30$^{\circ}$ should be robust to identifying galaxies with external interaction.

\bsp	
\label{lastpage}
\end{document}